\documentclass[aip,amsmath,amssymb,reprint]{revtex4-2}

\usepackage{graphicx}
\usepackage{dcolumn}
\usepackage{bm}
\usepackage[utf8]{inputenc}
\usepackage[T1]{fontenc}
\usepackage{mathptmx}
\usepackage{bm}
\usepackage{bbm}
\usepackage{amsfonts}
\usepackage{amsmath}
\usepackage{natbib}
\usepackage{dsfont} 
\providecommand{\keywords}[1]{\textbf{\textit{Index terms---}} #1}
\usepackage[mathlines]{lineno}
\usepackage[colorinlistoftodos]{todonotes}
\usepackage[colorlinks=true, allcolors=blue]{hyperref}
\usepackage{soul,xcolor}
\usepackage{siunitx}
\usepackage{braket}
\usepackage{textcomp}
\setstcolor{red}
\RequirePackage{color}

\begin{document}

\title[Waveguiding in massive two-dimensional Dirac systems]{Waveguiding in massive two-dimensional Dirac systems}
\author{V. G. Ibarra-Sierra}%
\email{ibarrasierra@uabc.edu.mx}
\affiliation{Facultad de Ciencias, Universidad Autónoma de Baja California, Apartado postal 1880, 22800 Ensenada, Baja California, México}%
\author{E. J. Robles-Raygoza}
 \affiliation{Facultad de Ciencias, Universidad Autónoma de Baja California, Apartado postal 1880, 22800 Ensenada, Baja California, México}%
\author{J. C. Sandoval-Santana}%
\affiliation{Centro de Nanociencias y Nanotecnolog\'ia, Universidad Nacional Aut\'onoma de M\'exico, Apartado Postal 2681, 22800 Ensenada, Baja California, M\'exico.}%
\affiliation{\'Area de F\'isica Te\'orica y Materia Condensada,
Universidad Aut\'onoma Metropolitana Azcapotzalco, Av. San Pablo 180,
Col. Reynosa-Tamaulipas, 02200 Cuidad de M\'exico, M\'exico.}
\author{R. Carrillo-Bastos}
\affiliation{Facultad de Ciencias, Universidad Autónoma de Baja California, Apartado postal 1880, 22800 Ensenada, Baja California, México}%

\preprint{AIP/123-QED}

\date{\today}
             
\begin{abstract}
The study of waveguide propagating modes is essential for achieving directional electronic transport in two-dimensional materials. Simultaneously, exploring potential gaps in these systems is crucial for developing devices akin to those employed in conventional electronics. Building upon the theoretical groundwork laid by Hartmann et al. \cite{Hartmann2014Waveguides}, which focused on implementing waveguides in pristine graphene monolayers, this work delves into the impact of a waveguide on two-dimensional gapped Dirac systems. We derive exact solutions encompassing wave functions and energy-bound states for a secant-hyperbolic attractive potential in gapped graphene, with a gap generated by sublattice asymmetry or a Kekulé-distortion. These solutions leverage the inherent properties and boundary conditions of the Heun polynomials. Our findings demonstrate that the manipulation of the number of accessible energy-bound states, i.e., transverse propagating modes, relies on factors such as the width and depth of the potential as well as the gap value of the two-dimensional material.

\end{abstract}
\maketitle
\section{Introduction}

\begin{figure}[t]
\includegraphics[width=0.28\textwidth]{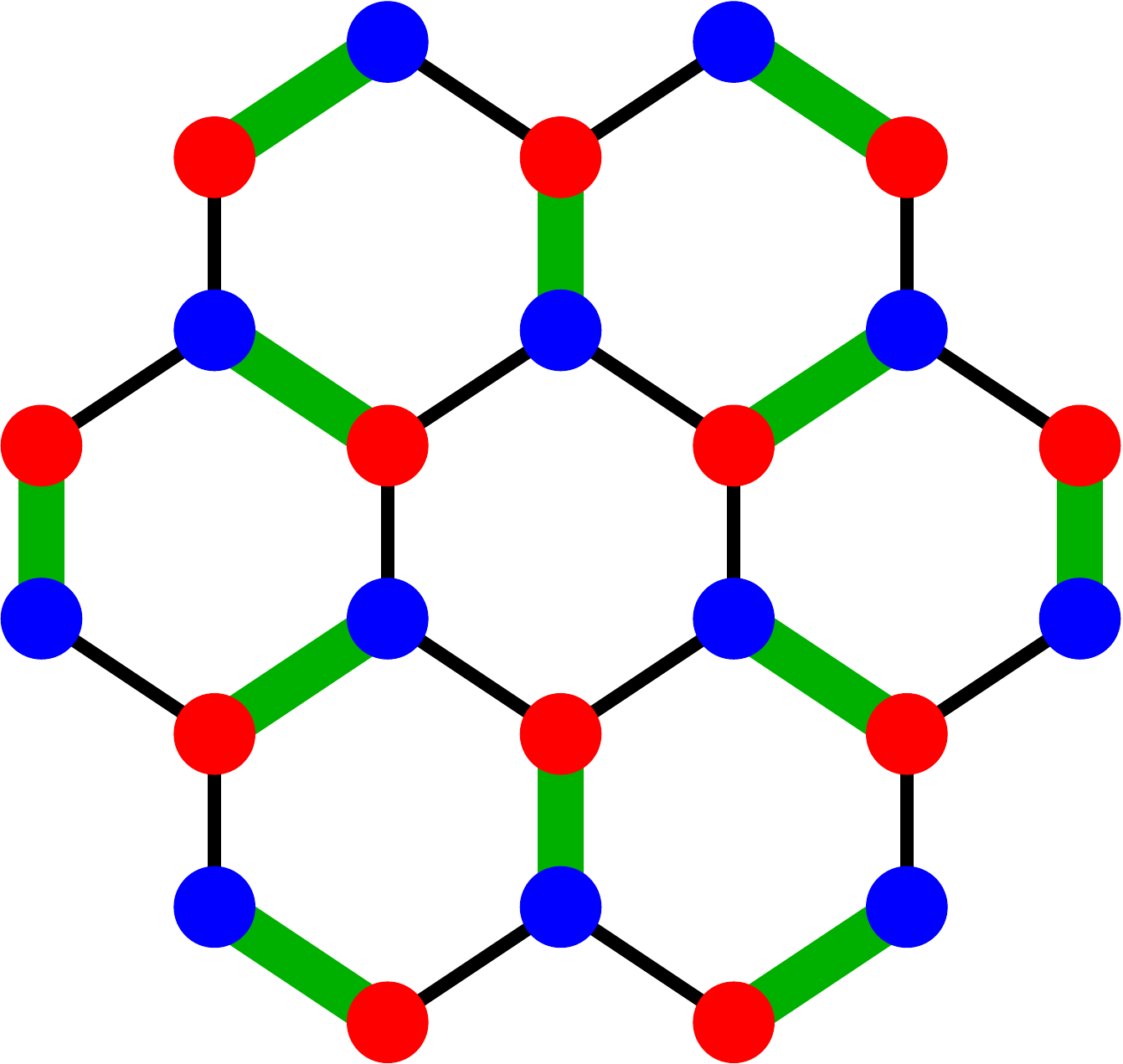}
\includegraphics[width=0.43\textwidth]{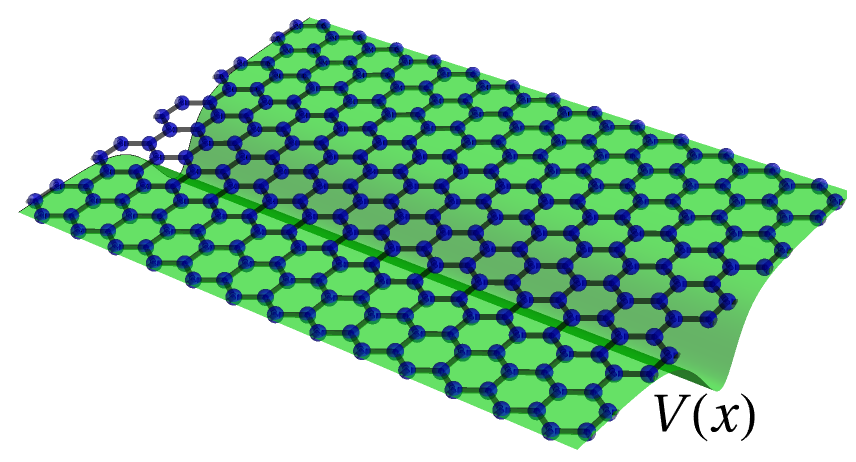}
\put(-110,108){\large(a)}
\put(-110,1){\large(b)}
\caption{\label{Fig:Graphene_KekO} Schematic diagrams of (a) Kekul\'e distorted honeycomb lattice with O-shaped texture and (b) graphene monolayer under hyperbolic-secant well potential defined by the Eq. (\ref{Eq:Potential_Secanthiperbolic})}
\end{figure}

Two-dimensional (2D) materials\cite{Xu2013,miro2014atlas,Mounet2018,Ibarra2019}, such as graphene, hBN, MoS$_2$, black phosphorus, and borophene ($8-Pmmn$), offer remarkable versatility in manipulating their electronic properties through surface modifications\cite{substrate01,substrate02,surface01}. This route presents exciting prospects for exploring novel quantum states and advancing technological applications\cite{Fiori2014,Xia2014,Lemme2014,khan2020recent}.  The ability to induce a band gap is crucial for creating devices comparable to those used in conventional electronics\cite{chaves2020bandgap}. In the case of graphene, numerous chemical and physical approaches have been proposed to achieve such a gap\cite{xu2018interfacial}. Notably, one primary method involves forming chemical bonds with substrates\cite{Varchon2007,Shemella2009}, which introduces an energy difference between the two sublattices of graphene, resulting in the emergence of an energy gap\cite{Ando2015}. An alternative approach is the intercalation of graphene with Li or Ca, which induces a periodic modulation of the graphene lattice [see Fig. \ref{Fig:Graphene_KekO}(a)], known as Kekulé-distorted graphene O-shaped (Kek-O)\cite{Gamayun2018, Mojarro2020-KekO,Bao2021, Bao2022,EOM2022, Qu2022,KekuleAndrade2022,Zeng2022}. There are other Kekulé distortions\cite{Gamayun2018,Stegmann2022}, but we will consider only the O-type here and use the generic term to refer to it. Theoretical calculations employing the tight-binding approximation have demonstrated that this distortion induces an effective gap in the electronic spectrum\cite{Gamayun2018,KekuleAndrade2022}.

In 2D materials, electronic transport can become quasi-one-dimensional by defining waveguides, allowing directional control of currents\cite{Xia2014} in materials like graphene\cite{Hartmann2010Waveguides, Hartmann2014Waveguides, Hartmann2017, Hartmann2020Waveguides,Carrillo2014Gaussian, Cao2017Waveguides, Mosallanejad2018Waveguides, Raygoza2022} or borophene\cite{Hartmann2021}. Analogous to light transport in optical fibers\cite{Addanki2018Review, Dragic2018}, in electronic waveguides, transport occurs through energy-bound states or propagation modes that depend on the nature of the waveguide parameters, such as their shape, intensity, and width\cite{Hartmann2010Waveguides, Hartmann2014Waveguides, Hartmann2020Waveguides, Raygoza2022}. Concerning the implementation of these waveguides, they can consist of local gate voltages\cite{Hartmann2010Waveguides, Cao2017Waveguides, Mosallanejad2018Waveguides}, strain deformations\cite{Carrillo2014Gaussian, Carrillo2016strained, Giambastiani2022}, or the approximation of charged carbon nanotubes to the surface of these 2D materials\cite{Hartmann2020Waveguides}. Lastly, one of the latest studies on implementing these waveguides has shown the possibility of electronic transport even in the presence of edge and bulk disorders in these 2D systems\cite{Mucciolo2009, Raygoza2022}.

In this work, we are interested in studying waveguiding in gapped Dirac-like systems via scalar potentials with the main focus on manipulating the accessible number of transverse propagating modes.  In particular, we study the Hamiltonian at low energies for gapped graphene, either by sublattice symmetry or by a Kekulé distortion, under a hyperbolic-secant well potential [see Figs. \ref{Fig:Graphene_KekO}(a)-(b)]. To solve the Schr\"odinger for these systems, we use a set of transformations to decouple the pseudospin components of the wave function and generate a general second-order ordinary differential equation with complex coefficients. We show that these second-order differential equations in both systems are equivalent; thus, we can treat them simultaneously. Subsequently, using a set of dimensionless variables \cite{Hartmann2014Waveguides}, we find exact solutions for these differential equations based on the Heun polynomials\cite{Maier2007,Olver2010nist,Downing2013,Hartmann2014Waveguides}. Hence, thanks to the properties of these polynomials and their boundary conditions, we find accessible energy-bound states are determined by the depth and width of the waveguide and by the gap of the 2D system. Furthermore, we analyze the behavior of symmetric functions' real and imaginary parts for the accessible transverse propagating modes due to the waveguide. Finally, we show that the probability density (electron density) is invariant to the gap value for one allowed bound state.

We organize the paper as follows, first, in Sec. \ref{sec:ModelsGrapheneGap}, we introduce the low energy Hamiltonian for gapped graphene under a secant-hyperbolic potential well, where the gap arises from sublattice asymmetry or a Kekulé distortion. Then, we apply a couple of transformations to obtain a general second-order ordinary differential for the wave function's spinor components. Next, we introduce dimensionless variables in Sec. \ref{sec:Exact_Solutions} to find the exact solution. This solution relies on the Heun polynomials, and their properties allow us to find the expressions of the energy-bound states as a function of the gap (quantization problem). Then, in Sec. \ref{sec:Energy_WaveFunctions} we study the behavior of the energy-bound states, the real/imaginary parts of the wave functions and probability density. Finally, we summarize and conclude in Sec. \ref{sec:Conclusion}. 

\section{Models of gapped graphene under secant-hyperbolic well potential}\label{sec:ModelsGrapheneGap}
\subsection{Hamiltonian for sublattice asymmetric graphene}\label{sec:ModelGraphene}

Let's consider the Hamiltonian 2D of graphene in the $\boldsymbol{K}$ valley in the presence of a one-dimensional  potential $V(x)$ as
\begin{equation}\label{Eq:HamiltonianGraphene}
 H_\text{\tiny G}=v_F\left(p_x\sigma_x +p_y\sigma_y \right) + \Delta_\text{\tiny G}\sigma_z+V(x)\sigma_0,
\end{equation}
where $v_F$ is the Fermi velocity in the pristine graphene monolayer, $p_x$ and $p_y$ are the momentum operators, $\sigma_i$ with $i=x,\, y,\, z$ are the Pauli matrices, and $\sigma_0$ is an identity matrix acting in pseudospin degree of freedom. The second term in Eq. (\ref{Eq:HamiltonianGraphene}) describes the effect of the gap that comes from the energy difference between the two graphene sublattices. Similar to previous works\cite{Hartmann2010Waveguides,Hartmann2014Waveguides, Hartmann2021}, we study the waveguide effect of a hyperbolic-secant potential well $V(x)$ defined as 
\begin{equation}\label{Eq:Potential_Secanthiperbolic}
V(x)=- U_0\,\mathrm{sech}(\lambda x),   
\end{equation}
where $U_0=\hbar v_F V_0$ and $\lambda$ are the strength and width, respectively. The schematic diagram of the graphene monolayer under the potential defined by Eq. (\ref{Eq:Potential_Secanthiperbolic}) is shown in Fig. \ref{Fig:Graphene_KekO}(b).

We aim to obtain the analytic expressions of the energy spectrum and wave functions associated with the Hamiltonian (\ref{Eq:HamiltonianGraphene}). Therefore, we use the stationary Schr\"odinger equation in the form
\begin{equation}\label{Eq:StationarySchrodinger_Graphene}
\mathcal{H}_\text{\tiny G}\boldsymbol{\Psi}_\text{\tiny G}(x,y)=\mathcal{E}_\text{\tiny G}\boldsymbol{\Psi}\text{\tiny G}(x,y), 
\end{equation}
where $\mathcal{H}_\text{\tiny G}=H_\text{\tiny G}/\hbar v_F$, $\mathcal{E}_\text{\tiny G}=E_\text{\tiny G}/\hbar v_F$, and $\boldsymbol{\Psi}_\text{\tiny G}(x,y)=\left[\,\psi_A(x,y),\,\psi_B(x,y)\,\right]^\top$ is a two-component spinor with $A$ and $B$ label the two sublattices of graphene.

To find analytical solutions of Eq. (\ref{Eq:StationarySchrodinger_Graphene}), first we apply the following unitary transformation 
\begin{equation}\label{Eq:UnitaryTransformation1}
U_\text{\tiny G}=\frac{1}{\sqrt{2}}\left(\sigma_x+\sigma_z\right),
\end{equation}
($U^\dagger_\text{\tiny G}U_\text{\tiny G}=1$) that changes the Eq. (\ref{Eq:StationarySchrodinger_Graphene}) in
\begin{equation}\label{Eq:StationarySchrodinger_Graphene_T1}
 \tilde{\mathcal{H}}_{\text{\tiny G}_1}\boldsymbol{\tilde{\Psi}}_\text{\tiny G}(x,y)=\mathcal{E}_\text{\tiny G}\boldsymbol{\tilde{\Psi}}_\text{\tiny G}(x,y),
\end{equation}
where $\tilde{\mathcal{H}}_{\text{\tiny G}_1}=\hat{U}_\text{\tiny G}^{\dagger}\mathcal{H}_\text{\tiny G}\hat{U}_\text{\tiny G}$ and the transformed spinor is given by $\tilde{\boldsymbol{\Psi}}\text{\tiny G}(x,y)=U_\text{\tiny G}^{\dagger}\boldsymbol{\Psi}_\text{\tiny G}(x,y)=\left[\,\tilde{\psi}_1(x,y),\,\tilde{\psi}_2(x,y)\,\right]^\top$ with the spin components defined as $\tilde{\psi}_{1,\,2}(x,y)=[\psi_A(x,y)\pm\psi_B(x,y)]/\sqrt{2}$. The next step is to multiply both sides of Eq. (\ref{Eq:StationarySchrodinger_Graphene_T1}) by the Pauli matrix $\sigma_z$, resulting in the following generalized eigenvalue problem.
\begin{equation}\label{Eq:StationarySchrodinger_Graphene_T2}
\tilde{\mathcal{H}}_{\text{\tiny G}_2}\boldsymbol{\tilde{\Psi}}_\text{\tiny G}(x,y)=\mathbb{E}_\text{\tiny G}\boldsymbol{\tilde{\Psi}}_\text{\tiny G}(x,y),
\end{equation}
where the matrix $\mathbb{E}_\text{\tiny G}=\mathcal{E}_\text{\tiny G}\sigma_z $ and the transformed Hamiltonian $\tilde{\mathcal{H}}_{\text{\tiny G}_2} =\sigma_z\tilde{\mathcal{H}}_{\text{\tiny G}_1}$ has the form
\begin{equation}\label{Eq:HaltonianH2_Graphene}
\tilde{\mathcal{H}}_{\text{\tiny G}_2}=\tilde{p}_x \sigma_0 + i \left(\tilde{p}_y\sigma_x +\tilde{\Delta}_\text{\tiny G}\sigma_y\right)+ \tilde{V}(x)\sigma_z,    
\end{equation}
with $\tilde{p}_x=-i\partial_{x}$, $\tilde{p}_y=-i\partial_{y}$, $\tilde{\Delta}_\text{\tiny G}=\Delta_\text{\tiny G}/\hbar v_F$ and $\tilde{V}(x)=V(x)/v_F \hbar$.

According to Eq. (\ref{Eq:Potential_Secanthiperbolic}), the confining potential varies only in the $x$-component; therefore, the $y$ component of the moment can be taken as a good quantum number. Thus, the wave function can be expressed as
\cite{Hartmann2010Waveguides,Hartmann2014Waveguides}
\begin{equation}\label{Eq:WaveFunction_Graphene}
\boldsymbol{\tilde{\Psi}}_\text{\tiny G}(x,y)=\exp{\left(ik_y y\right)}\boldsymbol{\tilde{\psi}}_\text{\tiny G}(x),
\end{equation}
with $\boldsymbol{\tilde{\psi}}_\text{\tiny G}(x)=\left[\,\tilde{\psi}_1(x),\,\tilde{\psi}_2(x)\,\right]^\top$ and the momentum $k_y \in \mathbb{R}$ in order to have propagate modes along the $y$-direction. Subsequently, using the Eq. (\ref{Eq:WaveFunction_Graphene}) into Eq. (\ref{Eq:StationarySchrodinger_Graphene_T2}), we can obtain the following ordinary differential equation
\begin{equation}\label{Eq:DifferentialEquation_G}
\frac{d}{dx}\boldsymbol{\tilde{\psi}}_\text{\tiny G}(x)+i\mathbb{H}_\text{\tiny G}(x)\boldsymbol{\tilde{\psi}}_\text{\tiny G}(x)=0,   
\end{equation}
where the matrix function $\mathbb{H}_\text{\tiny G}(x)$ is defined by
\begin{equation}\label{Eq:Function_H(x)_Graphene}
\mathbb{H}_\text{\tiny G}(x)=\left[\tilde{V}(x)-\mathcal{E}_\text{\tiny G}\right]\sigma_z+i\left[k_y\sigma_x +\tilde{\Delta}_m \sigma_y\right].    
\end{equation}

Following the treatment, we derive the Eq. (\ref{Eq:DifferentialEquation_G}) once more to respect to the $x$ component, and using the same equation again; we can find the following second-order ordinary differential equation
\begin{equation}\label{Eq:SetOrd2_DifEquation_Graphene}
\frac{d^2}{dx^2}\boldsymbol{\tilde{\psi}}_\text{\tiny G}(x)+\mathbb{F}_\text{\tiny G}(x)\boldsymbol{\tilde{\psi}}_\text{\tiny G}(x)=0,  
\end{equation}
where the new matrix function $\mathbb{F}_\text{\tiny G}(x)$ is given by
\begin{multline}\label{Eq:FunctionF_Graphene}
\mathbb{F}_\text{\tiny G}(x)=\left[i\frac{d}{dx}\tilde{V}(x)\right]\sigma_z\\+\left[\left(\tilde{V}(x)-\mathcal{E}_{G}\right)^2-\left(k_y^2+\tilde{\Delta}_{m}^2\right)\right]\sigma_0.
\end{multline}
From the Eqs. (\ref{Eq:SetOrd2_DifEquation_Graphene}) and (\ref{Eq:FunctionF_Graphene}), we can write the differential equation for each component as\cite{Hartmann2014Waveguides}
\begin{multline}\label{Eq:ComponetDifEq_Graphene}
\frac{d^2}{dx^2}\tilde{\psi}_{\nu}(x)+\left[i\nu\frac{d}{dx}\tilde{V}(x)\right.\\ 
+\left(\tilde{V}(x)-\mathcal{E}_\text{\tiny G}\right)^2
\left.
-|\tilde{k}_\text{\tiny G}|^2\right]\tilde{\psi}_{\nu}(x)=0,   
\end{multline}
where $|\tilde{k}_\text{\tiny G}|^2=k_y^2+\tilde{\Delta}_\text{\tiny G}^2$ and $\nu=\pm 1$ with $\tilde{\psi}_{+1}(x)=\tilde{\psi}_1(x)$ and $\tilde{\psi}_{-1}(x)=\tilde{\psi}_2(x)$. Compared with previous models\cite{Hartmann2010Waveguides,Hartmann2014Waveguides}, this second-order differential equation incorporates a modification of the wave number $k_y$ due to $\tilde{\Delta}_\text{\tiny G}$, representing the gap in this system. Subsequently, we demonstrate that this effect determines the allowed values of $k_y$ for obtaining exact solutions and the number of accessible energy-bound states in this system. We present the solution to Eq. (\ref{Eq:ComponetDifEq_Graphene}) in section \ref{sec:Energy_WaveFunctions}.

\subsection{Hamiltonian for Kekulé-distorted graphene}\label{sec:ModelKek0}
In this part, we study a monolayer of Kekulé-distorted graphene\cite{Gamayun2018,Mojarro2020-KekO,Bao2022,Qu2022} under the effect of the hyperbolic-secant potential well. The schematic diagram of the Kekulé-O texture is shown in Fig. \ref{Fig:Graphene_KekO}(a). The Hamiltonian for this system\cite{Beenakker2018} is given by
\begin{equation}\label{Eq:Hamiltonian_KO}
 H_{\text{\tiny KO}}=\tau_0\otimes\left[v_F\boldsymbol{\sigma}\cdot\boldsymbol{p}\right]+\left[\Delta_{\text{\tiny KO}}\tau_x\right]\otimes\sigma_z+\tau_0\otimes\left[V(x)\sigma_0\right]
\end{equation}
again, $\boldsymbol{p}=(p_x,p_y)$ and $\boldsymbol{\sigma}=(\sigma_x,\sigma_y)$ represent the momentum and Pauli operator vectors, respectively. The Pauli matrices $\sigma_x$, $\sigma_y$, $\sigma_z$, and the unit matrix $\sigma_0$ act on the pseudospin degree of freedom. On the other hand, the matrices $\tau_x$, $\tau_y$, $\tau_z$, and $\tau_0$ describe the valley degree of freedom. The potential $V(x)$, as defined in Eq. (\ref{Eq:Hamiltonian_KO}), is given by the expression (\ref{Eq:Potential_Secanthiperbolic}). Lastly, the final Hamiltonian can be conveniently expressed using the Kronecker product \cite{Harold1981Kronecker, Loan2000Kronecker}, which is defined as
\begin{equation}
 \mathcal{C}\otimes \mathcal{D}=  
  \begin{pmatrix}
c_{11} \mathcal{D} &  \cdots & c_{1n} \mathcal{D} \\
\vdots & \ddots  &  \vdots  \\
c_{m1} \mathcal{D}  & \cdots  & c_{nm} \mathcal{D} \\
\end{pmatrix}\nonumber.
\end{equation}
where $\mathcal{C}$ is a $m\times n$ matrix and $\mathcal{D}$ is $p\times q$ matrix.

Similar to the case of massive graphene described in Sec. \ref{sec:ModelGraphene}, we use a set of transformations to find a general decoupled differential equation. Hence, we start with the stationary Schr\"odinger equation as
\begin{equation}\label{Eq:StationarySchrodinger_KO}
\mathcal{H}_{\text{\tiny KO}}\boldsymbol{\Psi}_{\text{\tiny KO}}(x,y)=\mathcal{E}_{\text{\tiny KO}}\boldsymbol{\Psi}_{\text{\tiny KO}}(x,y), 
\end{equation}
with $\mathcal{H}_{\text{\tiny KO}}=H_{\text{\tiny KO}}/\hbar v_F$, $\mathcal{E}_{\text{\tiny KO}}=E_{\text{\tiny KO}}/\hbar v_F$ and  $\boldsymbol{\Psi}{\text{\tiny KO}}(x,y)=\exp{\left(ik_y y\right)}\left[\psi_{1}(x),\,\psi_{2}(x), \psi_{3}(x),\,\psi_{4}(x)\,\right]^\top$
is a four-component spinor, where again, $k_y$ is a good quantum number that describes the propagating modes in the $y$ component. The components of this spinor are the amplitudes on sublattices $A$ and $B$ for the valleys $K'$ and $K$\cite{Gamayun2018,Mojarro2020-KekO}, respectively. Now, we use the unitary transformation $U_{\text{\tiny KO}}=\tau_0\otimes \left(\sigma_x+\sigma_z\right)/\sqrt{2}$  and the multiplication by the matrix $\tau_0 \otimes \sigma_{z}$. Hence, these operations only affect the pseudospin space and leave the valley degree of freedom invariant. Thus,  Eq. (\ref{Eq:StationarySchrodinger_KO}) changes into the next generalized eigenvalue problem
\begin{equation}\label{Eq:Trasform2_KO}
\tilde{\mathcal{H}}_{{\text{\tiny KO}}}\boldsymbol{\tilde{\psi}}_{\text{\tiny KO}}(x)=\mathbb{E}{\text{\tiny KO}}\boldsymbol{\tilde{\psi}}_{\text{\tiny KO}}(x),  \end{equation}
where the matrix $\mathbb{E}_{\text{\tiny KO}}=\mathcal{E}_{\text{\tiny KO}}\left(\tau_0 \otimes\sigma_{z}\right)$ and the transform Hamiltonian $\tilde{\mathcal{H}}_{{\text{\tiny KO}}}=\left(\tau_0 \otimes \sigma_{z}\right)\left[U^\dagger_{\text{\tiny KO}}\mathcal{H}_{\text{\tiny KO}}U_{\text{\tiny KO}}\right]$ has the explicit form
\begin{multline}\label{Eq:Hamitonian_Tras_KO_1}
\tilde{\mathcal{H}}_{{\text{\tiny KO}}}=\tilde{p}_x \left(\tau_0 \otimes \sigma_{0}\right)+ +i\tilde{\Delta}_{\text{\tiny KO}}\left(\tau_x \otimes \sigma_{y}\right)
\\
+ ik_y\left(\tau_0 \otimes \sigma_{x}\right)+\tilde{V}(x) \left(\tau_0 \otimes \tau_{z}\right),   
\end{multline}
with $\tilde{\Delta}_{\text{\tiny KO}}=\Delta_{\text{\tiny KO}}/\hbar v_F$ and $\tilde{V}(x)=V(x)/\hbar v_{F}$. The spinor $\boldsymbol{\tilde{\psi}}_{\text{\tiny KO}}(x)=U^{\dagger}_{\text{\tiny KO}}\boldsymbol{\tilde{\psi}}_{\text{\tiny KO}}(x)$ in Eq. (\ref{Eq:Trasform2_KO}) is defined as
\begin{equation}
\boldsymbol{\tilde{\psi}}_{\text{\tiny KO}}(x)\\=\left[\,\tilde{\psi}_1(x),\,\tilde{\psi}_2(x),\, \tilde{\psi}_3(x), \, \tilde{\psi}_4(x)\right]^\top,    
\end{equation}
with $\tilde{\psi}_{i,\, i+1}(x)=[\psi_{i}(x)\pm\psi_{i+1}(x)]/\sqrt{2}$ and $i=1,\, 3$.

Subsequently, using Eqs. (\ref{Eq:Trasform2_KO}) and (\ref{Eq:Hamitonian_Tras_KO_1}), we can obtain the following differential equation
\begin{equation}\label{Eq:DifferentialEquation_KO}
\frac{d}{dx}\boldsymbol{\tilde{\psi}}_\text{\tiny KO}(x)+i\mathbb{H}_\text{\tiny KO}(x)\boldsymbol{\tilde{\psi}}_\text{\tiny KO}(x)=0,  
\end{equation}
where the function matrix $\mathbb{H}_\text{\tiny KO}(x)=\left[\tilde{V}(x)-\mathcal{E}_\text{\tiny G}\right] \left(\tau_0 \otimes\sigma_{z}\right)+ik_y\left(\tau_0\otimes \sigma_{x}\right)+i\tilde{\Delta}_{\text{\tiny KO}}\left(\tau_x \otimes \sigma_{y}\right)$. Similar to Sec. \ref{sec:ModelGraphene}, the second-order differential equation obtained from Eq. (\ref{Eq:DifferentialEquation_KO}) is given by
\begin{equation}\label{Eq:SecondDifEq_KekO}
\frac{d^2}{dx^2}{\tilde{\psi}}_{\text{\tiny KO}}(x)+\mathbb{F}_{\text{\tiny KO}}(x){\tilde{\psi}}_{\text{\tiny KO}}(x)=0, 
\end{equation} 
 where the matrix function $\mathbb{F}_{\text{\tiny KO}}(x)$ is given by
\begin{multline}\label{Eq:FunctionF_KekO}
\mathbb{F}_{\text{\tiny KO}}(x)=\left[i\frac{d}{dx}\tilde{V}(x)\right]\left(\tau_0\otimes \sigma_{z}\right)\\
+\left[\left(\tilde{V}(x)-\mathcal{E}_{\text{\tiny KO}}\right)^2-\left(k_y^2+\tilde{\Delta}_{\text{\tiny KO}}^2\right)\right]\left(\tau_0 \otimes\sigma_{0}\right).
\end{multline}

Once again, we can write each component of the previous second-order differential equation in a single component as follows 
\begin{multline}\label{Eq:ComponetDifEq_KekO}
\frac{d^2}{dx^2}\tilde{\psi}_{\nu}(x)+\left[i\nu\frac{d}{dx}\tilde{V}(x)\right.\\+
\left. \left(\tilde{V}(x)-\mathcal{E}_{\text{\tiny KO}}\right)^2
-|\tilde{k}_\text{\tiny KO}|^2 \right]\tilde{\psi}_{\nu}(x)=0, 
\end{multline}
where $|\tilde{k}_\text{\tiny KO}|^2=k_y^2+\tilde{\Delta}^2_\text{\tiny KO}$. Here, the index $\nu=\pm 1$ refers the spinor components $\tilde{\psi}_{+1}(x)=\tilde{\psi}_{1}(x)=\tilde{\psi}_{3}(x)$ and  $\tilde{\psi}_{-1}(x)=\tilde{\psi}_{2}(x)=\tilde{\psi}_{4}(x)$. This differential equation (\ref{Eq:ComponetDifEq_KekO}) has the same structure as the massive graphene (\ref{Eq:ComponetDifEq_Graphene}). We present its solution in the next section.

\section{Solutions and symmetric wave functions}\label{sec:Exact_Solutions}
\subsection{Exact Solutions}

To solve the differential equation in the case of massive graphene, whether generated by sublattice asymmetry (\ref{Eq:ComponetDifEq_Graphene}) or by a Kekulé distortion (\ref{Eq:ComponetDifEq_KekO}), we employ the method described in Ref. [\onlinecite{Hartmann2014Waveguides}]. Therefore, the second-order differential equation for the spinor components $\tilde{\psi}_{\nu}(x)$ in both systems can be expressed as
\begin{multline}\label{Eq:GeneralDiferentialEq}
\frac{d^2}{dx^2}\tilde{\psi}_{\nu}(x)+\left[i\nu\frac{d}{dx}\tilde{V}(x)\right.\\
\left.+ \left(\tilde{V}(x)-\mathcal{E}_{\text{\tiny (G, KO)}}\right)^2
-|\tilde{k}_{\text{\tiny (G, KO)}}|^2 \right]\tilde{\psi}_{\nu}(x)=0,   
\end{multline}
with $|\tilde{k}_{\text{\tiny (G, KO)}}|^2=k_y^2+\tilde{\Delta}^2_{\text{\tiny (G, KO)}}$. The Eq. (\ref{Eq:GeneralDiferentialEq}) can be conveniently written using the following definitions $\mathcal{E}_{\text{\tiny (G, KO)}}=\lambda\varepsilon$, $\tilde{V}(x)=-\omega\lambda\, \mathrm{sech}(\lambda x)$, $\omega=V_0/\lambda$, $k_y=\lambda\Gamma$, $\tilde{\Delta}_{\text{\tiny (G, KO)}}=\lambda\Gamma_g$ and $|\tilde{k}_{\text{\tiny (G, KO)}}|=\lambda |\tilde{\Gamma}|$; where $\varepsilon$, $\omega$,  $\Gamma$, $\Gamma_g$ and $|\tilde{\Gamma}|=\sqrt{\Gamma^2+\Gamma_{g}^2}$ are dimensionless variables. Under these definitions the Eq. (\ref{Eq:GeneralDiferentialEq}) takes the form
\begin{multline}\label{Eq:GeneralDiferentialEq_2}
\frac{d^2}{dx^2}\tilde{\psi}_{\nu}(x)+\lambda^2\left\{i\nu\frac{d}{dx}\left[\omega\,\mathrm{sech}(\lambda x)\right]\right.\\
\left.+\left[\omega\,\mathrm{sech}(\lambda x)-\varepsilon\right]^2-\tilde{\Gamma}^2 \right\}\tilde{\psi}_{\nu}(x)=0.   
\end{multline}

In analogy with the differential equation for the harmonic oscillator problem in nonrelativistic quantum mechanics\cite{shankar2012principles}, the exact solution of the above differential equation will allow us to find the analytical expressions of the wave functions and the energy spectrum for the massive graphene. Thus the exact solutions\cite{Hartmann2014Waveguides} of Eq. (\ref{Eq:GeneralDiferentialEq_2}) is given by
\begin{multline}\label{Eq:EigenFunctions_Solutions}
\tilde\psi_{\nu; n,m}(x)=A_{\nu}\left[\tilde{V}(x)\right]^{\kappa_{\nu; n}}\left[\zeta(x)-\frac{1}{2}\right]^{\mu_{\nu; n}}H_{\nu; n,m}(x),
\end{multline}
where $n=0,\,1,\,2,\, 3,...,$ and $m\leq n$ denote the number of propagating modes in the waveguide and $A_{\nu}$ is a constant of normalization. The parameters $\kappa_{\nu;n}$ and $\mu_{\nu;n}$ in the above expression are defined as
\begin{equation}\label{Eq:Kappa_Parameters}
\kappa_{\nu; n}=\omega-\frac{n+1}{2},  
\end{equation}
\vspace{-0.6cm}
\begin{equation}\label{Eq:Mu_Parameters}
\mu_{\nu; n}=-\frac{\nu}{2}-\frac{n}{2}.
\end{equation}
The function  $H_{\nu;\,n,m}(x)$ in Eq. (\ref{Eq:EigenFunctions_Solutions}) are the Heun polynomials\cite{Maier2007,Olver2010nist,Hartmann2014Waveguides} whose definition are 
\begin{multline}\label{Eq:HeunPolynomial_1}
H_{\nu; n,m}(x)=
H\left[a,\,q_{\nu; n, m};\,\alpha_{\nu; n},\,\beta_{\nu; n},\,\gamma_{\nu; n},\,\delta_{\nu; n};\,\zeta(x)\right].  
\end{multline}
where 
\begin{equation}\label{Eq:Zeta_x}
\zeta(x)=\frac{e^{-i\pi/4}}{\sqrt{2}}\frac{\tanh(\lambda x/2)+1}{\tanh(\lambda x/2)-i},
\end{equation} 
$a$ is a singularity parameter; $\alpha_{\nu; n},\,\beta_{\nu; n},\, \gamma_{\nu; n}, \, \delta_{\nu; n}$ are exponent parameters and $q_{\nu; n, m}$ is an accessory parameter\cite{Olver2010nist}. To generate these Heun polynomials, the parameters that define $H_{\nu; n,m}(x)$ should be defined as
\begin{equation}\label{Eq:a_Parameter}
a=\frac{1}{2},
\end{equation}
\vspace{-0.5cm}
\begin{equation}\label{Eq:Alfa_Parameters}
\alpha_{\nu ; n}=-n, 
\end{equation}
\vspace{-0.5cm}
\begin{equation}\label{Eq:q_Parameters}
q_{\nu;n,m}=\pm2i\omega\sqrt{|\tilde{\Gamma}_{n,m}|^2-\kappa_{\nu; n}^2}-n\kappa_{\nu;n}-\frac{n}{2}(1-\nu),
\end{equation}
\vspace{-0.5cm}
\begin{equation}\label{Eq:Beta_Parameter}
\beta_{\nu; n}=2\omega-n -\nu,
\end{equation}
\vspace{-0.5cm}
\begin{equation}\label{Eq:Gamma_Delta_Parameters}
\gamma_{\nu; n}=\delta_{\nu; n}=2\omega-n, 
\end{equation}
where $|\tilde{\Gamma}_{n,m}|^2=\Gamma^2_{n,m}+\Gamma_g^2$. Here, it is important to remark that due to the conditions of the Heun polynomials and the solutions for the wave functions, discrete values are found for the dimensionless parameter $\Gamma$. The expression for the Dirac spectrum (energy bound-states) in this system is given by\cite{Hartmann2014Waveguides}
\begin{equation}\label{Eq:Def_Enm}
\varepsilon_{n,m}=\pm\sqrt{|\tilde{\Gamma}_{n,m}|^2-\kappa_{\nu; n}^2}\,.   
\end{equation}
The values of $q_{\nu; n,m}$, with $\nu=\pm 1$, in Eq. (\ref{Eq:q_Parameters}) are determined from the eigenvalue problem 
\begin{equation}\label{Eq:EigenProblem_qnm}
 det(\mathbb{T}_{\nu; n}-q_{n,m}\mathbb{I})=0,   
\end{equation}
where $\mathbb{I}$ is a $(n+1)\times (n+1)$ unit matrix and  $\mathbb{T}_{\nu; n}$ is a tridiagonal matrix\cite{Hartmann2010Waveguides,Olver2010nist} defined as
\begin{equation}\label{eq:Definition_TMatrix}
\mathbb{T}_{\nu; n}=
\begin{pmatrix}
-Q_0^{\nu; n} & R_0^{\nu; n}  & 0  & \cdots & 0 \\
P_1^{\nu;n}  & -Q_1^{\nu;n} & R_1^{\nu;n} & \cdots & 0\\
0 & P_2^{\nu; n} & -Q_2^{\nu; n} & \cdots &  \vdots\\
\vdots & \vdots  &  \vdots  &  \ddots & R_{n-1}^{\nu; n}\\
0 & 0  & \cdots & P_n^{\nu; n} & -Q_n^{\nu; n}\\
\end{pmatrix},
\end{equation}
with $R_j^{\nu; n}=(j+1)(j+\gamma_{\nu; n})/2$, $P_j^{\nu; n}=(j-1+\alpha_{\nu; n})(j-1+\beta_{\nu; n})$ and $Q_{j}^{\nu; n}=j[3(j-1+\gamma_{\nu; n})/2+\delta_{\nu; n}/2-2\omega]$. The matrix $\mathbb{T}_{\nu; n}$ in Eq. (\ref{eq:Definition_TMatrix}) arises from the recursion relation of the solution of the Heun differential equation through the Fuchs-Frobenius method\cite{Olver2010nist}. Therefore, from the Eq. (\ref{Eq:EigenProblem_qnm}), the subindices in $q_{n,m}$ refer to the $m$-th eigenvalue associated with the $n$-th state. 

Now, using the Eqs.  (\ref{Eq:q_Parameters}), (\ref{Eq:Def_Enm}) and (\ref{Eq:EigenProblem_qnm}), it is possible to find the explicit form of the discrete expressions for $\Gamma_{n,m}$ and latter of energy bound-states $\varepsilon_{n,m}$  as a function of $\omega$ and the dimensionless gap $\Gamma_g$. In particular, the discrete expressions of $\Gamma_{n,m}$ can take negative and positive values for a fixed value of $\omega$ and $\Gamma_g$. In Appendix \ref{Appendix_One}, we show the analytic expressions for $\Gamma_{n,m}$ and $\varepsilon_{n,m}$ only in the cases where $n=0,\,1\,,2$ and $m\leq n$. 

The boundary conditions for Eq. (\ref{Eq:EigenFunctions_Solutions}) are $\lim_{ x\to \pm \infty} \tilde{\psi}_{\nu; n, m}(x)= 0$ for each energy-bound state. Therefore, from the  properties of  Heun polynomials\cite{Maier2007,Hartmann2014Waveguides},  we have  $\lim_{ x\to \pm \infty} H_{\nu; n, m}(x)= 1$ and the term $\lim_{ x\to \pm \infty} \left[\zeta(x)-1/2\right]^{\mu_{\nu;n}}=(-1/2)^{\mu_{\nu;n}}$ where the exponent ${\mu_{\nu;n}}$ can take negatives and positives values. Lastly, the term $\lim_{ x\to \pm \infty}\left[\tilde{V}(x)\right]^{\kappa_{\nu; n}}=0$ only if $\kappa_{\nu; n}>0$. From this last condition and Eq. (\ref{Eq:Kappa_Parameters}), we find the following rule for the possibles number of energy-bound states
\begin{equation}\label{Eq:Condition_Discrete values}
 0 \leq n<2\omega-1,
\end{equation}
and hence, the lowest state $n=0$ for $\Gamma_g=0$ indicates the minimal relationship between the depth and width of secant-hyperbolic potential well to find one of the energy-bound states, that is, $V_0>\lambda/2$. 

Otherwise, from the Eq. (\ref{Eq:DifferentialEquation_G}) or (\ref{Eq:DifferentialEquation_KO}) is important to note that the connection between the functions $\tilde{\psi}_{+1}(x)$ and $\tilde{\psi}_{-1}(x)$ in Eq. (\ref{Eq:EigenFunctions_Solutions}) are given by
\begin{multline}\label{Eq:EigenFunction_plus}
\tilde{\psi}_{+1;n,m}(x)\\=\frac{ie^{-i\theta_{n,m}}}{|\tilde{\Gamma}_{m,n}|}\left[\varepsilon_{n,m}-\frac{1}{\lambda}\left(\tilde{V}(x)+i\frac{d}{dx}\right)\right]\tilde{\psi}_{-1,n,m}(x), 
\end{multline}
and
\begin{multline}\label{Eq:EigenFunction_minus}
\tilde{\psi}_{-1;n,m}(x)\\=-\frac{ie^{i\theta_{n,m}}}{|\tilde{\Gamma}_{m,n}|}\left[\varepsilon_{n,m}-\frac{1}{\lambda}\left(\tilde{V}(x)-i\frac{d}{dx}\right)\right]\tilde{\psi}_{+1,n,m}(x), 
\end{multline}
where $\tan\theta_{n,m}=\Gamma_g/\Gamma_{n,m}$.
Therefore, it is possible to determine only one of the following solutions, for example, $\tilde{\psi}_{+1,n,m}(x)$ and the solution of $\tilde{\psi}_{-1,n,m}(x)$ is determined by Eq. (\ref{Eq:EigenFunction_minus}).

\subsection{Symmetric wave functions}
Again in analogy with the nonrelativistic quantum harmonic oscillator problem, the wave functions have a definite parity, i.e., odd and even functions for each quantum level. In this sense, the wave functions $\tilde{\psi}_{+1,n,m}(x)$ and $\tilde{\psi}_{-1,n,m}(x)$ are not the symmetric wave functions associated to hyperbolic-secant potential well (\ref{Eq:Potential_Secanthiperbolic}). To find these symmetric wave functions, we introduce the following transformation 
\begin{equation}\label{Eq:OddEven_Functions}
\begin{bmatrix}
\psi_{\,\mathrm{I}; n, m}(x)/A_\mathrm{I}\\
\psi_{\,\mathrm{II}; n, m}(x)/A_\mathrm{II}\\
\end{bmatrix}=
U_{\text{\tiny SC}}
\begin{bmatrix}
\tilde{\psi}_{+1,n,m}(x)\\
\tilde{\psi}_{-1,n,m}(x)
\end{bmatrix},
\end{equation} 
where $A_\mathrm{I}$ and $A_\mathrm{II}$ are two normalization constants and
\begin{equation}\label{Eq:Symmetric_UnitaryTrasformation}
U_{\text{\tiny SC}}=\frac{1}{\sqrt{2}}
\begin{bmatrix} 
e^{i\Theta^{+}_{n,m}} & e^{-i\Theta^{+}_{n,m}}  \\
e^{i\Theta^{-}_{n,m}} & e^{-i\Theta^{-}_{n,m}}   \\
\end{bmatrix},
\end{equation} 
is another unitary transformation  
with $\Theta^{\pm}_{n,m}=\theta_{n,m}/2\pm \left(\pi/ 4\right)=\arctan\left(\Gamma_g/\Gamma_{n,m}\right)/2\pm\left(\pi/4\right)$.  Using Eq. (\ref{Eq:Symmetric_UnitaryTrasformation}) the Eqs. (\ref{Eq:EigenFunction_plus}) and (\ref{Eq:EigenFunction_minus}) can be rewrite as
\begin{multline}\label{Eq:First_DifSymmetrizedSystem}
 \left[\tilde{V}(x)+\lambda|\tilde{\Gamma}_{n, m}|-\lambda\varepsilon_{n,m}\right]\left[\frac{1}{A_\mathrm{II}}\psi_{\,\mathrm{II}, n, m}(x)\right]\\
 -\frac{d}{dx}\left[\frac{1}{A_\mathrm{I}}\psi_{\,\mathrm{I}, n, m}(x)\right]=0,   
\end{multline}
\begin{multline}\label{Eq:Second_DifSymmetrizedSystem}
 \left[\tilde{V}(x)-\lambda|\tilde{\Gamma}_{n, m}|-\lambda\varepsilon_{n,m}\right]\left[\frac{1}{A_\mathrm{I}}\psi_{\,\mathrm{I}, n, m}(x)\right]\\
 +\frac{d}{dx}\left[\frac{1}{A_\mathrm{II}}\psi_{\,\mathrm{II}, n, m}(x)\right]=0.   
\end{multline}
From the above system of differential equations, we can note that $\psi_{\,\mathrm{I}, n, m}(x)$ and $\psi_{\,\mathrm{II}, n, m}(x)$ keep a defined parity, for example, if we change $x\rightarrow\, -x$ in Eqs. (\ref{Eq:First_DifSymmetrizedSystem})-(\ref{Eq:Second_DifSymmetrizedSystem}) and if it is satisfied that $\psi_{\,\mathrm{I}, n, m}(-x)=\psi_{\,\mathrm{I}, n, m}(x)$,  $\psi_{\,\mathrm{II}, n, m}(-x)= -\psi_{\,\mathrm{II}, n, m}(x)$, hence this system of differential equations remain invariant under potential parity $\tilde{V}(-x)=\tilde{V}(x)$.

\begin{figure}[t]
\centering
\includegraphics[width=0.415\textwidth]{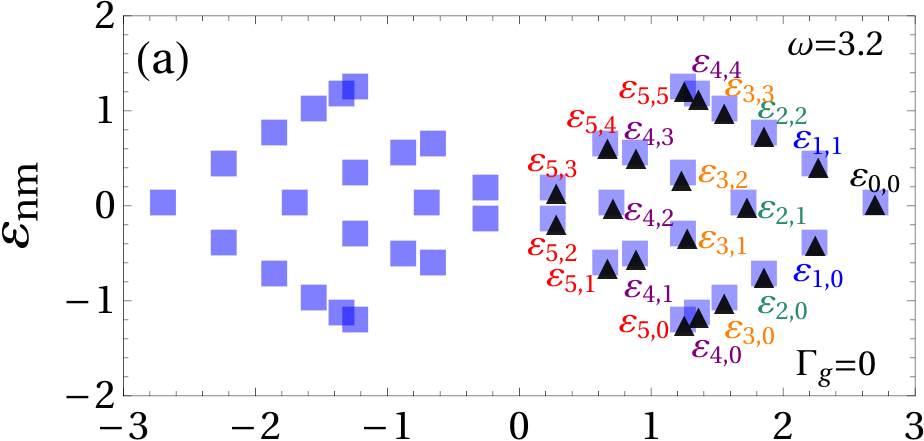}
\includegraphics[width=0.415\textwidth]{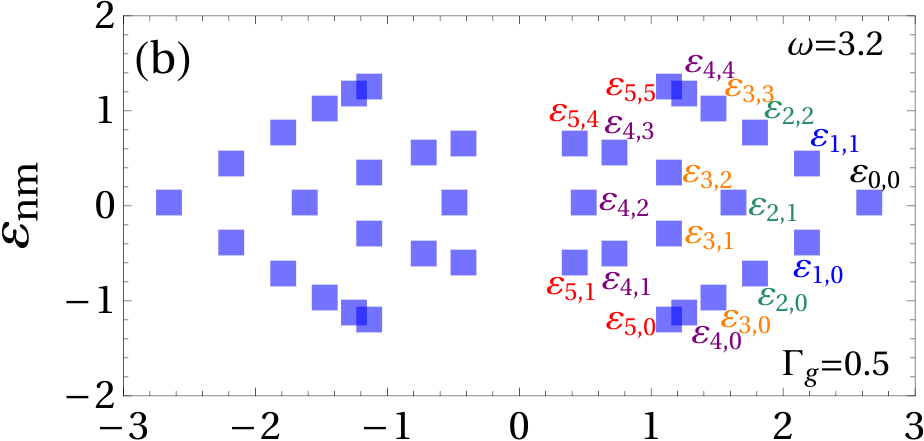}
\includegraphics[width=0.415\textwidth]{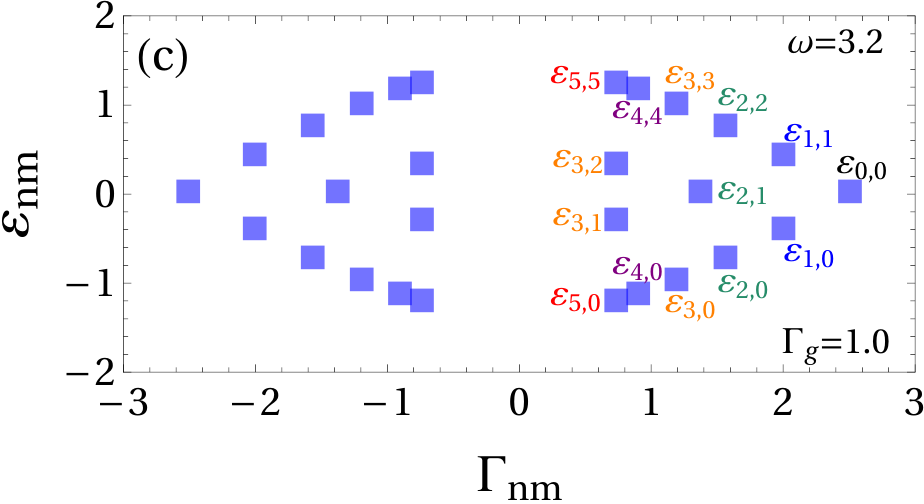}
\includegraphics[width=0.4155\textwidth]{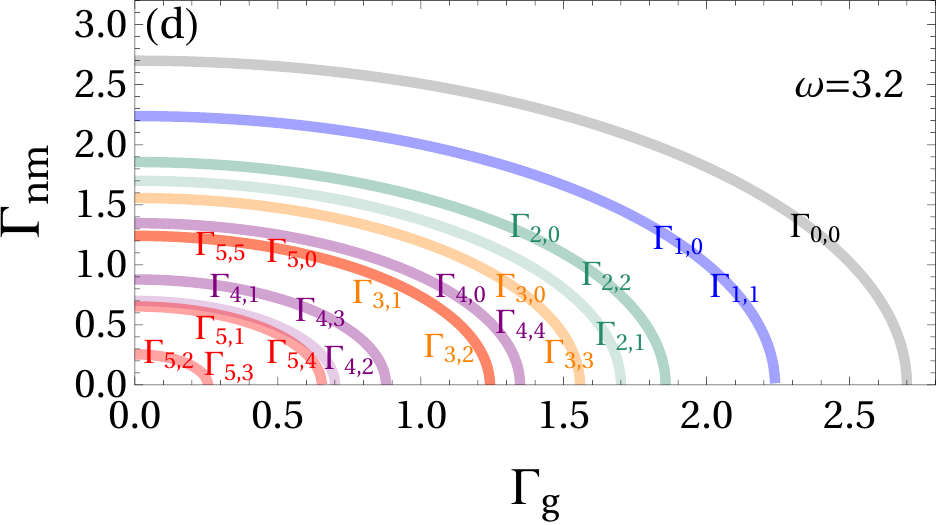}
\caption{\label{fig:EnergySpectrum} Dirac energy spectrum $\varepsilon_{n,m}$ and discrete values $\Gamma_{n,m}$. In panels (a), (b), and (c), we show the Dirac energy spectrum (\ref{Eq:Def_Enm}) (blue square points) as a function of discrete values $\Gamma_{n,m}$ for $n=0, 1, 2, 3, 4, 5$,\, $m=0, 1, 2, ..., n\,$, $\omega=3.2$ and using the dimensionless gap values: $\Gamma_g=0$, $\Gamma_g=0.5$, $\Gamma_g=1$, respectively. In panel (a), the black triangles correspond to the values of $\varepsilon_{n,m}$ calculated in the Ref. [\onlinecite{Hartmann2014Waveguides}]. In panel (d), we show the behavior of the discrete values $\Gamma_{n,m}>0$ as a function of dimensionless gap $\Gamma_g$.
}
\end{figure}

\section{Energy spectrum and wave functions}\label{sec:Energy_WaveFunctions}

\begin{figure}[t]
\centering
\includegraphics[width=0.17\textwidth]{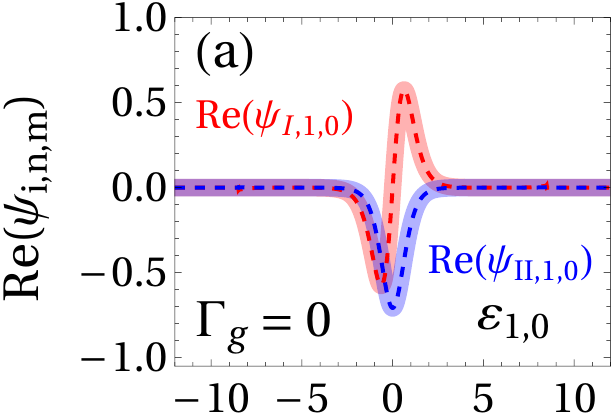}
\includegraphics[width=0.15\textwidth]{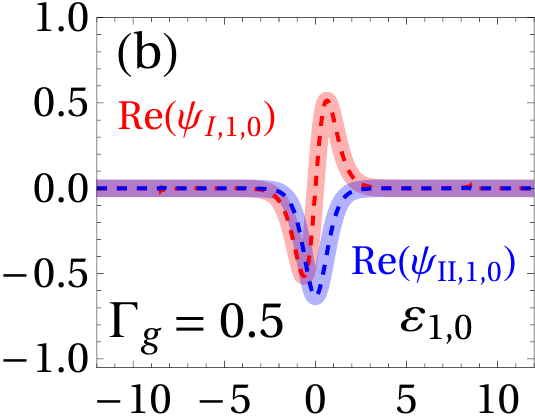}
\includegraphics[width=0.15\textwidth]{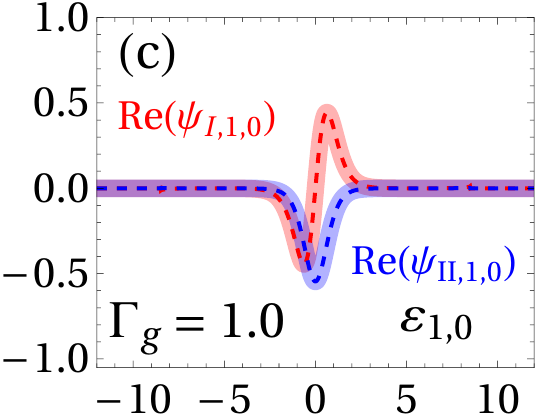}
\includegraphics[width=0.17\textwidth]{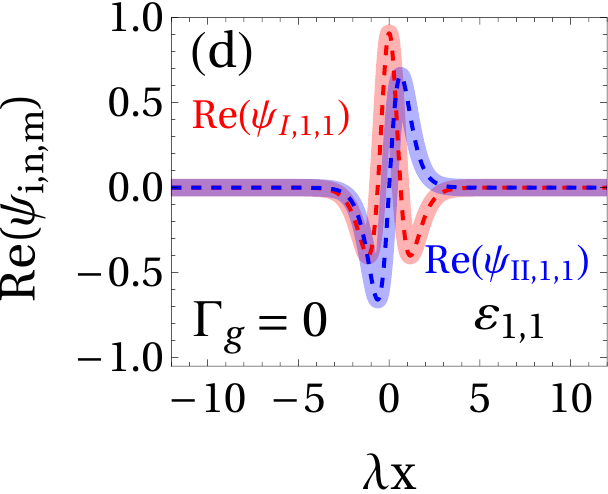}
\includegraphics[width=0.15\textwidth]{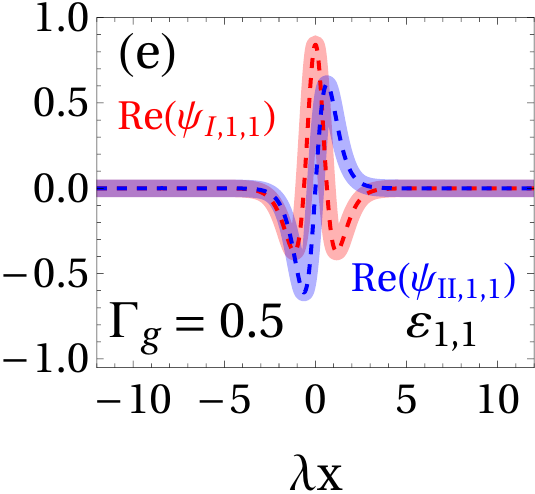}
\includegraphics[width=0.15\textwidth]{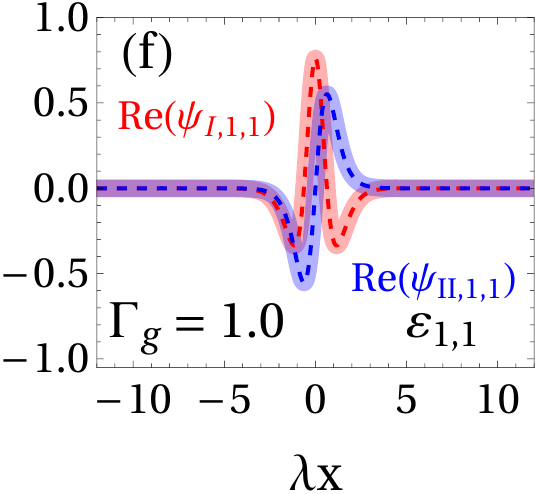}
\caption{\label{fig:ReWaveFunctions} Real part of the wave functions Re$[\psi_{\,\mathrm{i}, n, m}(x)]$ with $i=\,$I (red) and $i=\,$II (blue) using $\omega=3.2$, $\Gamma_{n,m}>0$ and for the dimensionless gap values $\Gamma_g=0$, $\Gamma_g=0.5$ and $\Gamma_g=1$. In panels (a), (b), and (c), we show Re$[\psi_{\,\mathrm{i}, n, m}(x)]$ for $\Gamma_{1,0}$ and $\varepsilon_{1,0}$. The panels (d), (e), and (f) are the Re$[\psi_{\,\mathrm{i}, n, m}(x)]$ for $\Gamma_{1,1}$ and $\varepsilon_{1,1}$. In all these panels, the light solid lines correspond to Re$[\psi_{\,\mathrm{i}, n, m}(x)]$ given by (\ref{Eq:OddEven_Functions}), and the dashed lines denote the corresponding numerical solution of Eqs. (\ref{Eq:First_DifSymmetrizedSystem}) and (\ref{Eq:Second_DifSymmetrizedSystem}).
}
\end{figure}
\begin{figure}[t]
\centering
\includegraphics[width=0.17\textwidth]{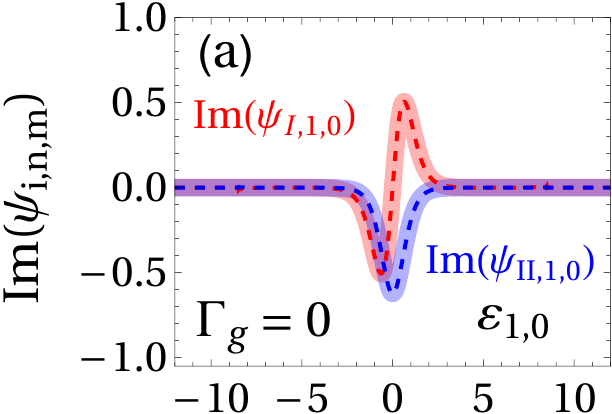}
\includegraphics[width=0.15\textwidth]{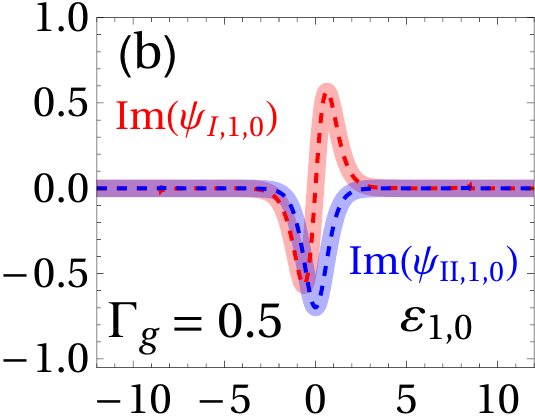}
\includegraphics[width=0.15\textwidth]{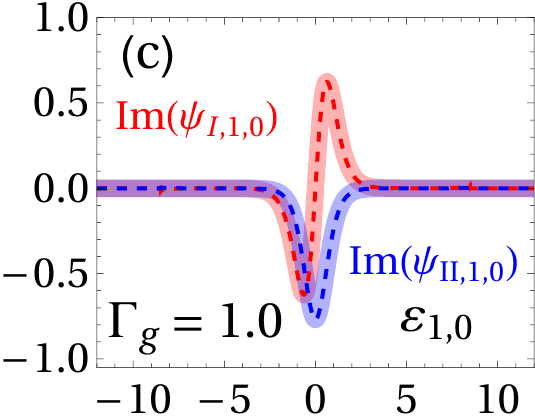}
\includegraphics[width=0.17\textwidth]{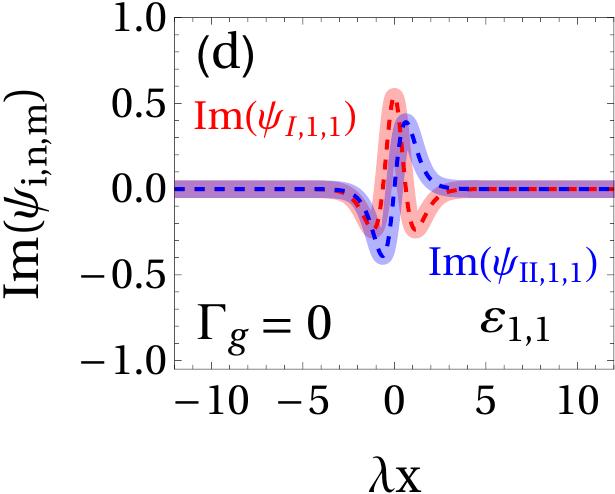}
\includegraphics[width=0.15\textwidth]{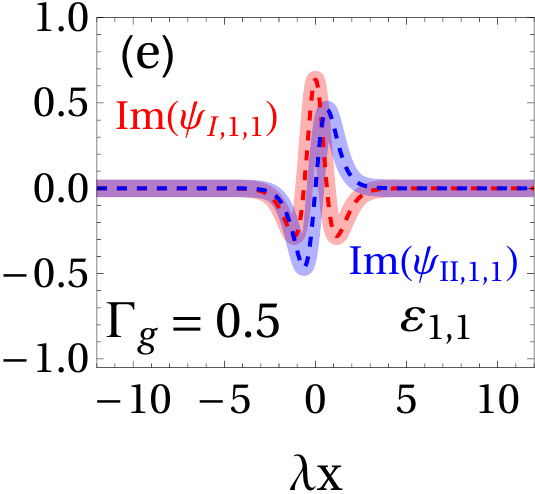}
\includegraphics[width=0.15\textwidth]{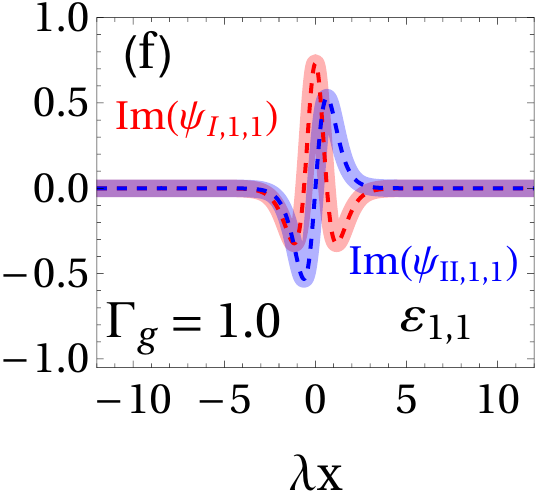}
\caption{\label{fig:ImWaveFunctions} Imaginary part of the wave functions Im$[\psi_{\,\mathrm{i}, n, m}(x)]$ with $i=\,$I (red) and $i=\,$II (blue) using $\omega=3.2$, $\Gamma_{n,m}>0$ and for the dimensionless gap values $\Gamma_g=0$, $\Gamma_g=0.5$ and $\Gamma_g=1$. In panels (a), (b), and (c), we show Im$[\psi_{\,\mathrm{i}, n, m}(x)]$ for $\Gamma_{1,0}$ and $\varepsilon_{1,0}$. The panels (d), (e), and (f) are the Im$[\psi_{\,\mathrm{i}, n, m}(x)]$ for $\Gamma_{1,1}$ and $\varepsilon_{1,1}$. In all these panels, the light solid lines correspond to Im$[\psi_{\,\mathrm{i}, n, m}(x)]$ given by (\ref{Eq:OddEven_Functions}), and the dashed lines denote the corresponding numerical solution of Eqs. (\ref{Eq:First_DifSymmetrizedSystem}) and (\ref{Eq:Second_DifSymmetrizedSystem}).
}
\end{figure}

We analyze the effects of the hyperbolic-secant potential well in 2D gapped systems discussed in the previous section. To obtain the energy spectrum and wave functions from exact solutions, we use that relation between the strength and width of the hyperbolic-secant well potential has a specific value\cite{Hartmann2014Waveguides} $\omega=3.2$. Therefore using the Eq. (\ref{Eq:Condition_Discrete values}), we found that $n<5.4$ and the possibles discrete values are $n=0,\,1,\,2,\,3,\,4,\,5$ and $m\leq n$ ($m=0,\,1,...,\,n$).

In Figs. \ref{fig:EnergySpectrum}(a), \ref{fig:EnergySpectrum}(b) and \ref{fig:EnergySpectrum}(c), we show the Dirac energy spectrum (blue square points) as a function of discrete values $\Gamma_{n,m}$ and for the values of dimensionless gap $\Gamma_g=0$, $\Gamma_g=0.5$, and $\Gamma_g=1$. Charge carriers in these  systems are relativistic fermions; this makes a difference when studying the effect of the hyperbolic-secant potential well and its energy-bound states. One of the main differences is that these bound states exist for positive and negative symmetric energies\cite{Hartmann2014Waveguides}.  Therefore, in Fig. \ref{fig:EnergySpectrum}(a), the blue square points are the energy spectrum for $\Gamma_g=0$. This spectrum is symmetric and degenerate for the discrete values $\Gamma_{n,m}<0$ and $\Gamma_{n,m}>0$. In this same figure, we compare the same spectrum (black triangle points) concerning the obtained in the Ref. [\onlinecite{Hartmann2014Waveguides}] for massless graphene ($\Gamma_g=0$) and the case of exact solutions. The Figs. \ref{fig:EnergySpectrum}(b) and \ref{fig:EnergySpectrum}(c) show the energy spectrum for $\Gamma_g=0.5$, and $\Gamma_g=1$, respectively. These figures show two main effects on the increase in the band gap. First, there are forbidden values of $\Gamma_{n,m}$ and fewer accessible energy states, which can be interpreted as manipulation of propagating modes in the waveguide. Second, the energy states present a small translation towards zero of $\Gamma_{m,n}$ maintaining its value for a fixed value $\omega=3.2$. Third, we found that there are protected bound states for $-\omega\leq\epsilon_{n,m}\leq0$ and unprotected  bound states $\epsilon_{n,m}>0$. This protection is given by the attractive potential. This is consistent with previous results in 
guided electrons against disorder in graphene nanoribbons
\cite{Raygoza2022}.

Regarding the behavior of $\Gamma_{n,m}$, in Fig \ref{fig:EnergySpectrum}(d), we show its conduct as a function of the dimensionless gap. Since that $\Gamma_{n,m}=\left[k_y/\lambda\right]_{n,m}$ and following Eq.  (\ref{Eq:WaveFunction_Graphene}) for massive graphene, $\Gamma_{n,m}$ must be a real number. However, these states with $\omega=3.2$ have a real value which subsequently tends to zero when $\Gamma_g$ increases. To a higher value of $\Gamma_g$, $\Gamma_{n,m}$  becomes a complex number. This transition from a real to imaginary value in $\Gamma_{n,m}$ generates forbidden states in the energy spectrum. Therefore, this provides a straightforward way to manipulate the number of propagating states.

In the panels of Fig. (\ref{fig:ReWaveFunctions}) and Fig. (\ref{fig:ImWaveFunctions}), we plot the real and imaginary parts of the wave functions $\psi_{,\mathrm{I}; n, m}(x)$ (blue lines) and $\psi_{,\mathrm{II}; n, m}(x)$ (red lines). These panels illustrate the cases where: $\Gamma_{n,m}>0$; the dimensionless gaps are $\Gamma_g=0$, $\Gamma_g=0.5$, and $\Gamma_g=1$; and the states are $\varepsilon_{1,0}$ and $\varepsilon_{1,1}$. We also include a comparison between the exact solutions given by Eqs. (\ref{Eq:EigenFunctions_Solutions}) and (\ref{Eq:OddEven_Functions}) (blue/red light solid lines) and the corresponding numerical solutions obtained from Eqs. (\ref{Eq:First_DifSymmetrizedSystem})-(\ref{Eq:Second_DifSymmetrizedSystem}) (blue/red dashed lines).

From the behavior of the wave functions' real and imaginary parts, several characteristics can be observed. Firstly, they change their parity when the state $\varepsilon_{n,m}$ is changed. This behavior is analogous to the nonrelativistic case for wave functions in the quantum harmonic oscillator problem. Secondly, the cases with $\Gamma_g=0$ in Figs. \ref{fig:ReWaveFunctions}(a) and \ref{fig:ReWaveFunctions}(d) for the real part of the wave functions agree with the results reported in Ref. [\onlinecite{Hartmann2014Waveguides}].

Another characteristic is their behavior concerning the dimensionless gap. Thanks to the unitary transformation (\ref{Eq:Symmetric_UnitaryTrasformation}), it can be observed that they preserve their odd or even parity for different gap values. In the cases of $\Gamma_g=0.5$ and $\Gamma_g=1$, compared to $\Gamma_g=0$, the amplitude of the real part of these wave functions decreases while the imaginary part increases. This effect is due to Eq. (\ref{Eq:Symmetric_UnitaryTrasformation}), as changing the value of $\Gamma_g$ in the matrix elements $e^{\pm i\Theta^{\pm}{n,m}}$ modifies the real and imaginary parts of $\psi{,\mathrm{I}; n, m}(x)$ and $\psi_{,\mathrm{II}; n, m}(x)$.

\begin{figure}[t]
\centering
\includegraphics[width=0.1708\textwidth]{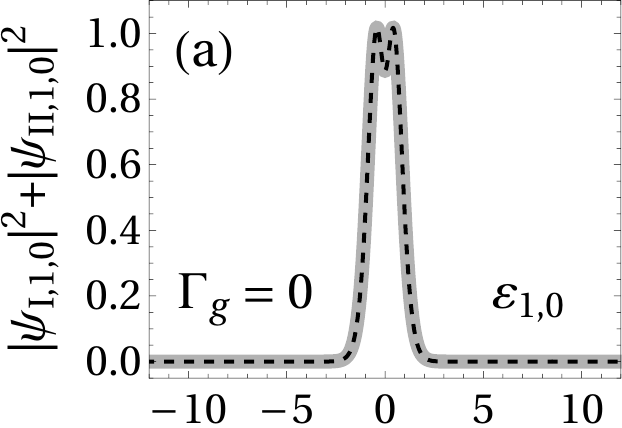}
\includegraphics[width=0.15\textwidth]{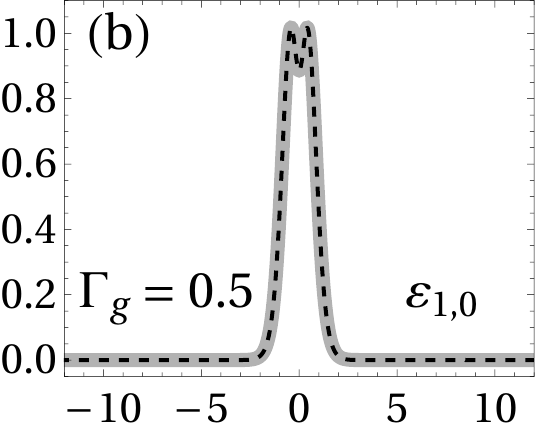}
\includegraphics[width=0.15\textwidth]{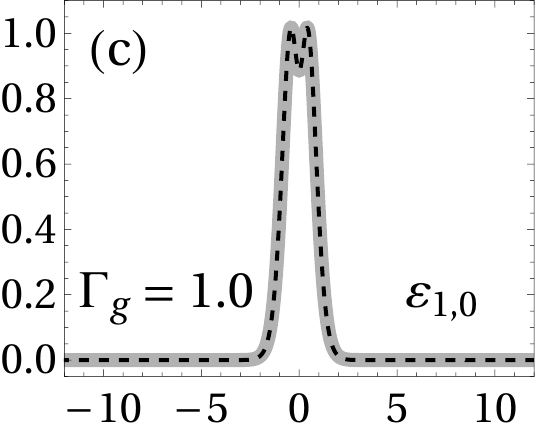}
\includegraphics[width=0.1708\textwidth]{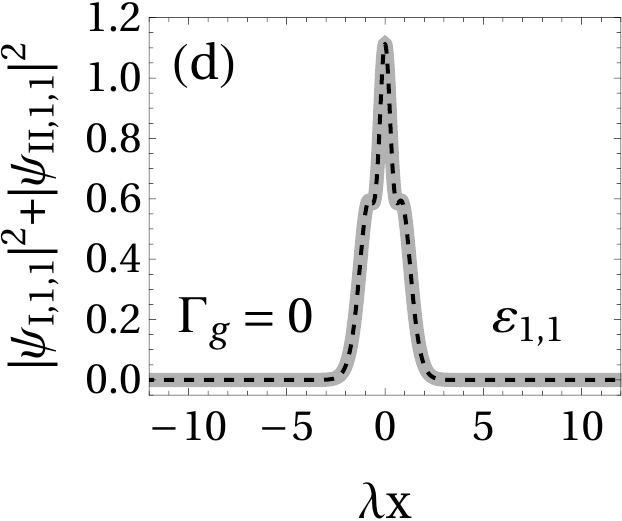}
\includegraphics[width=0.15\textwidth]{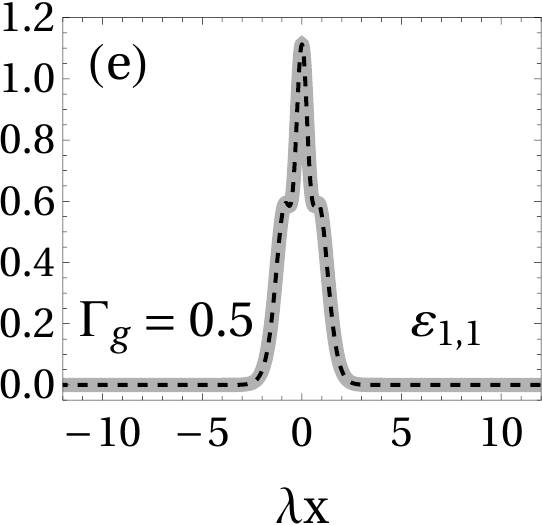}
\includegraphics[width=0.15\textwidth]{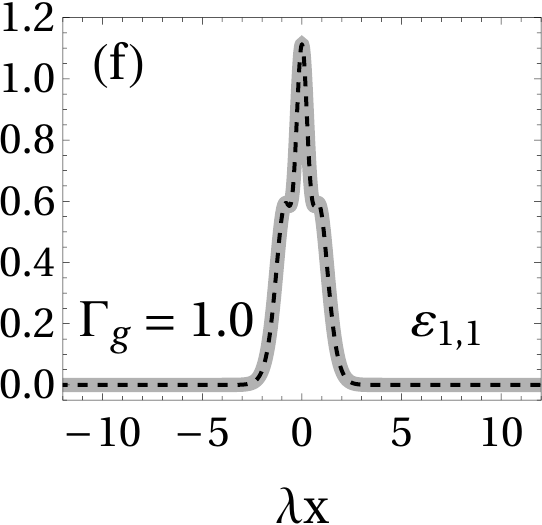}
\caption{\label{fig:ProbabilityDensity} Probability density $|\psi_{\,\mathrm{I}, n, m}(x)|^2+|\psi_{\,\mathrm{II}, n, m}(x)|^2$ for $\omega=3.2$, $\Gamma_{n,m}>0$ and using the dimensionless gap values: $\Gamma_g=0$, $\Gamma_g=0.5$ and $\Gamma_g=1.0$. In panels (a), (b), and (c), we show the probability density for $\Gamma_{1,0}$ and $\varepsilon_{1,0}$. The panels (d), (e), and (f) is the probability density for $\Gamma_{1,1}$ and $\varepsilon_{1,1}$. In all these panels, the light solid lines correspond to the
probability density obtained from the wave functions (\ref{Eq:OddEven_Functions}). The dashed lines denote the numerical solution using Eqs. (\ref{Eq:First_DifSymmetrizedSystem}) and (\ref{Eq:Second_DifSymmetrizedSystem}). From these results, it is clear that the probability density is invariant under to the value of $\Gamma_g$. 
}
\end{figure}

However, as shown in the panels of Fig. \ref{fig:ProbabilityDensity}, the probability density $|\psi_{,\mathrm{I}, n, m}(x)|^2+|\psi_{,\mathrm{II}, n, m}(x)|^2$ for $\varepsilon_{1,0}$ and $\varepsilon_{1,1}$ remains invariant regardless of the value of $\Gamma_g$. Once again, in Fig. \ref{fig:ProbabilityDensity}, the solid gray line corresponds to the probability density using the exact solutions of Eqs. (\ref{Eq:EigenFunctions_Solutions}) and (\ref{Eq:OddEven_Functions}), and the dotted lines are obtained from the numerical solution of Eqs. (\ref{Eq:First_DifSymmetrizedSystem}) and (\ref{Eq:Second_DifSymmetrizedSystem}).

\section{Conclusion}\label{sec:Conclusion}
We present the study of energy-bound states and wave functions in a two-dimensional gapped system in the presence of the secant-hyperbolic potential well. Following the case of graphene without a gap in Ref. [\onlinecite{Hartmann2014Waveguides}], we apply certain transformations in the stationary Schr\"odinger equation to find a decouple second-order differential equation for each spinor element of the wave function. Then, we find the exact solutions in these 2D systems using a set of dimensionless variables. These exact solutions are proportional to Heun polynomials, and their properties allow us to find the energy-bound states as a function of the gap of the 2D system. We show that the accessible number of propagating modes depends on the potential's width and depth and the value of the gap of the system. In particular, we study this behavior in a gapped graphene system  with the gap generated by sublattice asymmetry or
by a Kekulé distortion.

\section{Acknowledgements}\label{sec:Acknowledgements}
E.J.R.-R acknowledges financial support from CONACyT.
V.G.I.-S and J.C.S.-S. acknowledge the total support from Estancias Posdoctorales por M\'exico 2021 and 2022 CONACYT.

\appendix{}
\section{}\label{Appendix_One}

In this part, we show the list of discrete values $\Gamma_{n,m}$ and eigenvalues $\varepsilon_{n,m}$ as a function of dimensionless parameters $\omega$ and $\Gamma_g$ for $n=0,\,1,\,2$ and $m\leq n$ ($m=0,\,1,...,\,n$).
\begin{eqnarray}
\nonumber\\
\Gamma_{0,0}&=&\pm \frac{1}{2}\sqrt{1-4\Gamma_g^2 - 4\omega + 4\omega^2}\nonumber\\ 
\Gamma_{1,0}&=&\pm \frac{1}{2\sqrt{\omega}}\sqrt{4\omega^3-8\omega^2-4\Gamma_g^2\omega+5\omega-1}\nonumber\\
\Gamma_{1,1}&=&\pm \frac{1}{2\sqrt{\omega}}\sqrt{4\omega^3-8\omega^2-4\Gamma_g^2\omega+5\omega-1}\nonumber\\
\Gamma_{2,0}&=&\pm \frac{1}{2\omega}\sqrt{4\omega^4-12\omega^3-4\Gamma_g^2\omega^2+13\omega^2-6\omega+1}\nonumber\\
\Gamma_{2,1}&=&\pm \frac{1}{2}\sqrt{4\omega^2-12\omega-4\Gamma_g^2+9}\nonumber\\
\Gamma_{2,2}&=&\pm \frac{1}{2\omega}\sqrt{4\omega^4-12\omega^3-4\Gamma_g^2\omega^2+13\omega^2-6\omega+1}\nonumber\\
\varepsilon_{0,0}&=&0\nonumber\\
\varepsilon_{1,0}&=&-\frac{1}{2}\sqrt{\frac{\omega-1}{\omega}}\nonumber\\
\varepsilon_{1,1}&=&\frac{1}{2}\sqrt{\frac{\omega-1}{\omega}}\nonumber\\
\varepsilon_{2,0}&=&-\frac{1}{2\omega}\sqrt{4\omega^2-6\omega+1}\nonumber\\
\varepsilon_{2,1}&=&0 \nonumber\\
\varepsilon_{2,2}&=&\frac{1}{2\omega}\sqrt{4\omega^2-6\omega+1}\nonumber
\end{eqnarray}

\bibliography{references.bib}

\begin{thebibliography}{46}%
\makeatletter
\providecommand \@ifxundefined [1]{%
 \@ifx{#1\undefined}
}%
\providecommand \@ifnum [1]{%
 \ifnum #1\expandafter \@firstoftwo
 \else \expandafter \@secondoftwo
 \fi
}%
\providecommand \@ifx [1]{%
 \ifx #1\expandafter \@firstoftwo
 \else \expandafter \@secondoftwo
 \fi
}%
\providecommand \natexlab [1]{#1}%
\providecommand \enquote  [1]{``#1''}%
\providecommand \bibnamefont  [1]{#1}%
\providecommand \bibfnamefont [1]{#1}%
\providecommand \citenamefont [1]{#1}%
\providecommand \href@noop [0]{\@secondoftwo}%
\providecommand \href [0]{\begingroup \@sanitize@url \@href}%
\providecommand \@href[1]{\@@startlink{#1}\@@href}%
\providecommand \@@href[1]{\endgroup#1\@@endlink}%
\providecommand \@sanitize@url [0]{\catcode `\\12\catcode `\$12\catcode
  `\&12\catcode `\#12\catcode `\^12\catcode `\_12\catcode `\%12\relax}%
\providecommand \@@startlink[1]{}%
\providecommand \@@endlink[0]{}%
\providecommand \url  [0]{\begingroup\@sanitize@url \@url }%
\providecommand \@url [1]{\endgroup\@href {#1}{\urlprefix }}%
\providecommand \urlprefix  [0]{URL }%
\providecommand \Eprint [0]{\href }%
\providecommand \doibase [0]{https://doi.org/}%
\providecommand \selectlanguage [0]{\@gobble}%
\providecommand \bibinfo  [0]{\@secondoftwo}%
\providecommand \bibfield  [0]{\@secondoftwo}%
\providecommand \translation [1]{[#1]}%
\providecommand \BibitemOpen [0]{}%
\providecommand \bibitemStop [0]{}%
\providecommand \bibitemNoStop [0]{.\EOS\space}%
\providecommand \EOS [0]{\spacefactor3000\relax}%
\providecommand \BibitemShut  [1]{\csname bibitem#1\endcsname}%
\let\auto@bib@innerbib\@empty
\bibitem [{\citenamefont {Hartmann}\ and\ \citenamefont
  {Portnoi}(2014)}]{Hartmann2014Waveguides}%
  \BibitemOpen
  \bibfield  {author} {\bibinfo {author} {\bibfnamefont {R.~R.}\ \bibnamefont
  {Hartmann}}\ and\ \bibinfo {author} {\bibfnamefont {M.~E.}\ \bibnamefont
  {Portnoi}},\ }\bibfield  {title} {\enquote {\bibinfo {title} {Quasi-exact
  solution to the dirac equation for the hyperbolic-secant potential},}\ }\href
  {https://doi.org/10.1103/PhysRevA.89.012101} {\bibfield  {journal} {\bibinfo
  {journal} {Phys. Rev. A}\ }\textbf {\bibinfo {volume} {89}},\ \bibinfo
  {pages} {012101} (\bibinfo {year} {2014})}\BibitemShut {NoStop}%
\bibitem [{\citenamefont {Xu}\ \emph {et~al.}(2013)\citenamefont {Xu},
  \citenamefont {Liang}, \citenamefont {Shi},\ and\ \citenamefont
  {Chen}}]{Xu2013}%
  \BibitemOpen
  \bibfield  {author} {\bibinfo {author} {\bibfnamefont {M.}~\bibnamefont
  {Xu}}, \bibinfo {author} {\bibfnamefont {T.}~\bibnamefont {Liang}}, \bibinfo
  {author} {\bibfnamefont {M.}~\bibnamefont {Shi}},\ and\ \bibinfo {author}
  {\bibfnamefont {H.}~\bibnamefont {Chen}},\ }\bibfield  {title} {\enquote
  {\bibinfo {title} {Graphene-like two-dimensional materials},}\ }\href
  {https://doi.org/10.1021/cr300263a} {\bibfield  {journal} {\bibinfo
  {journal} {Chemical Reviews}\ }\textbf {\bibinfo {volume} {113}},\ \bibinfo
  {pages} {3766--3798} (\bibinfo {year} {2013})}\BibitemShut {NoStop}%
\bibitem [{\citenamefont {Mir{\'o}}, \citenamefont {Audiffred},\ and\
  \citenamefont {Heine}(2014)}]{miro2014atlas}%
  \BibitemOpen
  \bibfield  {author} {\bibinfo {author} {\bibfnamefont {P.}~\bibnamefont
  {Mir{\'o}}}, \bibinfo {author} {\bibfnamefont {M.}~\bibnamefont
  {Audiffred}},\ and\ \bibinfo {author} {\bibfnamefont {T.}~\bibnamefont
  {Heine}},\ }\bibfield  {title} {\enquote {\bibinfo {title} {An atlas of
  two-dimensional materials},}\ }\href@noop {} {\bibfield  {journal} {\bibinfo
  {journal} {Chemical Society Reviews}\ }\textbf {\bibinfo {volume} {43}},\
  \bibinfo {pages} {6537--6554} (\bibinfo {year} {2014})}\BibitemShut {NoStop}%
\bibitem [{\citenamefont {Mounet}\ \emph {et~al.}(2018)\citenamefont {Mounet},
  \citenamefont {Gibertini}, \citenamefont {Schwaller}, \citenamefont {Campi},
  \citenamefont {Merkys}, \citenamefont {Marrazzo}, \citenamefont {Sohier},
  \citenamefont {Castelli}, \citenamefont {Cepellotti}, \citenamefont {Pizzi},\
  and\ \citenamefont {Marzari}}]{Mounet2018}%
  \BibitemOpen
  \bibfield  {author} {\bibinfo {author} {\bibfnamefont {N.}~\bibnamefont
  {Mounet}}, \bibinfo {author} {\bibfnamefont {M.}~\bibnamefont {Gibertini}},
  \bibinfo {author} {\bibfnamefont {P.}~\bibnamefont {Schwaller}}, \bibinfo
  {author} {\bibfnamefont {D.}~\bibnamefont {Campi}}, \bibinfo {author}
  {\bibfnamefont {A.}~\bibnamefont {Merkys}}, \bibinfo {author} {\bibfnamefont
  {A.}~\bibnamefont {Marrazzo}}, \bibinfo {author} {\bibfnamefont
  {T.}~\bibnamefont {Sohier}}, \bibinfo {author} {\bibfnamefont {I.~E.}\
  \bibnamefont {Castelli}}, \bibinfo {author} {\bibfnamefont {A.}~\bibnamefont
  {Cepellotti}}, \bibinfo {author} {\bibfnamefont {G.}~\bibnamefont {Pizzi}},\
  and\ \bibinfo {author} {\bibfnamefont {N.}~\bibnamefont {Marzari}},\
  }\bibfield  {title} {\enquote {\bibinfo {title} {Two-dimensional materials
  from high-throughput computational exfoliation of experimentally known
  compounds},}\ }\href {https://doi.org/10.1038/s41565-017-0035-5} {\bibfield
  {journal} {\bibinfo  {journal} {Nature Nanotechnology}\ }\textbf {\bibinfo
  {volume} {13}},\ \bibinfo {pages} {246--252} (\bibinfo {year}
  {2018})}\BibitemShut {NoStop}%
\bibitem [{\citenamefont {Ibarra-Sierra}\ \emph {et~al.}(2019)\citenamefont
  {Ibarra-Sierra}, \citenamefont {Sandoval-Santana}, \citenamefont {Kunold},\
  and\ \citenamefont {Naumis}}]{Ibarra2019}%
  \BibitemOpen
  \bibfield  {author} {\bibinfo {author} {\bibfnamefont {V.~G.}\ \bibnamefont
  {Ibarra-Sierra}}, \bibinfo {author} {\bibfnamefont {J.~C.}\ \bibnamefont
  {Sandoval-Santana}}, \bibinfo {author} {\bibfnamefont {A.}~\bibnamefont
  {Kunold}},\ and\ \bibinfo {author} {\bibfnamefont {G.~G.}\ \bibnamefont
  {Naumis}},\ }\bibfield  {title} {\enquote {\bibinfo {title} {Dynamical band
  gap tuning in anisotropic tilted dirac semimetals by intense elliptically
  polarized normal illumination and its application to
  $8\text{\ensuremath{-}}pmmn$ borophene},}\ }\href
  {https://doi.org/10.1103/PhysRevB.100.125302} {\bibfield  {journal} {\bibinfo
   {journal} {Phys. Rev. B}\ }\textbf {\bibinfo {volume} {100}},\ \bibinfo
  {pages} {125302} (\bibinfo {year} {2019})}\BibitemShut {NoStop}%
\bibitem [{\citenamefont {Zhao}\ \emph {et~al.}(2023)\citenamefont {Zhao},
  \citenamefont {Guo}, \citenamefont {Li}, \citenamefont {Wang}, \citenamefont
  {Peng}, \citenamefont {Zhong}, \citenamefont {Chen}, \citenamefont {Yu},
  \citenamefont {Xu}, \citenamefont {Xie}, \citenamefont {Gao}, \citenamefont
  {Wang},\ and\ \citenamefont {Hu}}]{substrate01}%
  \BibitemOpen
  \bibfield  {author} {\bibinfo {author} {\bibfnamefont {T.}~\bibnamefont
  {Zhao}}, \bibinfo {author} {\bibfnamefont {J.}~\bibnamefont {Guo}}, \bibinfo
  {author} {\bibfnamefont {T.}~\bibnamefont {Li}}, \bibinfo {author}
  {\bibfnamefont {Z.}~\bibnamefont {Wang}}, \bibinfo {author} {\bibfnamefont
  {M.}~\bibnamefont {Peng}}, \bibinfo {author} {\bibfnamefont {F.}~\bibnamefont
  {Zhong}}, \bibinfo {author} {\bibfnamefont {Y.}~\bibnamefont {Chen}},
  \bibinfo {author} {\bibfnamefont {Y.}~\bibnamefont {Yu}}, \bibinfo {author}
  {\bibfnamefont {T.}~\bibnamefont {Xu}}, \bibinfo {author} {\bibfnamefont
  {R.}~\bibnamefont {Xie}}, \bibinfo {author} {\bibfnamefont {P.}~\bibnamefont
  {Gao}}, \bibinfo {author} {\bibfnamefont {X.}~\bibnamefont {Wang}},\ and\
  \bibinfo {author} {\bibfnamefont {W.}~\bibnamefont {Hu}},\ }\bibfield
  {title} {\enquote {\bibinfo {title} {Substrate engineering for wafer-scale
  two-dimensional material growth: strategies{,} mechanisms{,} and
  perspectives},}\ }\href {https://doi.org/10.1039/D2CS00657J} {\bibfield
  {journal} {\bibinfo  {journal} {Chem. Soc. Rev.}\ }\textbf {\bibinfo {volume}
  {52}},\ \bibinfo {pages} {1650--1671} (\bibinfo {year} {2023})}\BibitemShut
  {NoStop}%
\bibitem [{\citenamefont {Niu}, \citenamefont {Zhang},\ and\ \citenamefont
  {Chen}(2019)}]{substrate02}%
  \BibitemOpen
  \bibfield  {author} {\bibinfo {author} {\bibfnamefont {T.}~\bibnamefont
  {Niu}}, \bibinfo {author} {\bibfnamefont {J.}~\bibnamefont {Zhang}},\ and\
  \bibinfo {author} {\bibfnamefont {W.}~\bibnamefont {Chen}},\ }\bibfield
  {title} {\enquote {\bibinfo {title} {Surface engineering of two-dimensional
  materials},}\ }\href {https://doi.org/https://doi.org/10.1002/cnma.201800181}
  {\bibfield  {journal} {\bibinfo  {journal} {ChemNanoMat}\ }\textbf {\bibinfo
  {volume} {5}},\ \bibinfo {pages} {6--23} (\bibinfo {year} {2019})},\ \Eprint
  {https://arxiv.org/abs/https://onlinelibrary.wiley.com/doi/pdf/10.1002/cnma.201800181}
  {https://onlinelibrary.wiley.com/doi/pdf/10.1002/cnma.201800181} \BibitemShut
  {NoStop}%
\bibitem [{\citenamefont {Jiang}\ \emph {et~al.}(2019)\citenamefont {Jiang},
  \citenamefont {Xu}, \citenamefont {Lu}, \citenamefont {Sun},\ and\
  \citenamefont {Ni}}]{surface01}%
  \BibitemOpen
  \bibfield  {author} {\bibinfo {author} {\bibfnamefont {J.}~\bibnamefont
  {Jiang}}, \bibinfo {author} {\bibfnamefont {T.}~\bibnamefont {Xu}}, \bibinfo
  {author} {\bibfnamefont {J.}~\bibnamefont {Lu}}, \bibinfo {author}
  {\bibfnamefont {L.}~\bibnamefont {Sun}},\ and\ \bibinfo {author}
  {\bibfnamefont {Z.}~\bibnamefont {Ni}},\ }\bibfield  {title} {\enquote
  {\bibinfo {title} {Defect engineering in 2d materials: Precise manipulation
  and improved functionalities},}\ }\href
  {https://doi.org/10.34133/2019/4641739} {\bibfield  {journal} {\bibinfo
  {journal} {Research}\ }\textbf {\bibinfo {volume} {2019}} (\bibinfo {year}
  {2019}),\ 10.34133/2019/4641739},\ \Eprint
  {https://arxiv.org/abs/https://spj.science.org/doi/pdf/10.34133/2019/4641739}
  {https://spj.science.org/doi/pdf/10.34133/2019/4641739} \BibitemShut
  {NoStop}%
\bibitem [{\citenamefont {Fiori}\ \emph {et~al.}(2014)\citenamefont {Fiori},
  \citenamefont {Bonaccorso}, \citenamefont {Iannaccone}, \citenamefont
  {Palacios}, \citenamefont {Neumaier}, \citenamefont {Seabaugh}, \citenamefont
  {Banerjee},\ and\ \citenamefont {Colombo}}]{Fiori2014}%
  \BibitemOpen
  \bibfield  {author} {\bibinfo {author} {\bibfnamefont {G.}~\bibnamefont
  {Fiori}}, \bibinfo {author} {\bibfnamefont {F.}~\bibnamefont {Bonaccorso}},
  \bibinfo {author} {\bibfnamefont {G.}~\bibnamefont {Iannaccone}}, \bibinfo
  {author} {\bibfnamefont {T.}~\bibnamefont {Palacios}}, \bibinfo {author}
  {\bibfnamefont {D.}~\bibnamefont {Neumaier}}, \bibinfo {author}
  {\bibfnamefont {A.}~\bibnamefont {Seabaugh}}, \bibinfo {author}
  {\bibfnamefont {S.~K.}\ \bibnamefont {Banerjee}},\ and\ \bibinfo {author}
  {\bibfnamefont {L.}~\bibnamefont {Colombo}},\ }\bibfield  {title} {\enquote
  {\bibinfo {title} {Electronics based on two-dimensional materials},}\ }\href
  {https://doi.org/10.1038/nnano.2014.207} {\bibfield  {journal} {\bibinfo
  {journal} {Nature Nanotechnology}\ }\textbf {\bibinfo {volume} {9}},\
  \bibinfo {pages} {768--779} (\bibinfo {year} {2014})}\BibitemShut {NoStop}%
\bibitem [{\citenamefont {Xia}\ \emph {et~al.}(2014)\citenamefont {Xia},
  \citenamefont {Wang}, \citenamefont {Xiao}, \citenamefont {Dubey},\ and\
  \citenamefont {Ramasubramaniam}}]{Xia2014}%
  \BibitemOpen
  \bibfield  {author} {\bibinfo {author} {\bibfnamefont {F.}~\bibnamefont
  {Xia}}, \bibinfo {author} {\bibfnamefont {H.}~\bibnamefont {Wang}}, \bibinfo
  {author} {\bibfnamefont {D.}~\bibnamefont {Xiao}}, \bibinfo {author}
  {\bibfnamefont {M.}~\bibnamefont {Dubey}},\ and\ \bibinfo {author}
  {\bibfnamefont {A.}~\bibnamefont {Ramasubramaniam}},\ }\bibfield  {title}
  {\enquote {\bibinfo {title} {Two-dimensional material nanophotonics},}\
  }\href {https://doi.org/10.1038/nphoton.2014.271} {\bibfield  {journal}
  {\bibinfo  {journal} {Nature Photonics}\ }\textbf {\bibinfo {volume} {8}},\
  \bibinfo {pages} {899--907} (\bibinfo {year} {2014})}\BibitemShut {NoStop}%
\bibitem [{\citenamefont {Lemme}\ \emph {et~al.}(2014)\citenamefont {Lemme},
  \citenamefont {Li}, \citenamefont {Palacios},\ and\ \citenamefont
  {Schwierz}}]{Lemme2014}%
  \BibitemOpen
  \bibfield  {author} {\bibinfo {author} {\bibfnamefont {M.~C.}\ \bibnamefont
  {Lemme}}, \bibinfo {author} {\bibfnamefont {L.-J.}\ \bibnamefont {Li}},
  \bibinfo {author} {\bibfnamefont {T.}~\bibnamefont {Palacios}},\ and\
  \bibinfo {author} {\bibfnamefont {F.}~\bibnamefont {Schwierz}},\ }\bibfield
  {title} {\enquote {\bibinfo {title} {Two-dimensional materials for electronic
  applications},}\ }\href {https://doi.org/10.1557/mrs.2014.138} {\bibfield
  {journal} {\bibinfo  {journal} {MRS Bulletin}\ }\textbf {\bibinfo {volume}
  {39}},\ \bibinfo {pages} {711–718} (\bibinfo {year} {2014})}\BibitemShut
  {NoStop}%
\bibitem [{\citenamefont {Khan}\ \emph {et~al.}(2020)\citenamefont {Khan},
  \citenamefont {Tareen}, \citenamefont {Aslam}, \citenamefont {Wang},
  \citenamefont {Zhang}, \citenamefont {Mahmood}, \citenamefont {Ouyang},
  \citenamefont {Zhang},\ and\ \citenamefont {Guo}}]{khan2020recent}%
  \BibitemOpen
  \bibfield  {author} {\bibinfo {author} {\bibfnamefont {K.}~\bibnamefont
  {Khan}}, \bibinfo {author} {\bibfnamefont {A.~K.}\ \bibnamefont {Tareen}},
  \bibinfo {author} {\bibfnamefont {M.}~\bibnamefont {Aslam}}, \bibinfo
  {author} {\bibfnamefont {R.}~\bibnamefont {Wang}}, \bibinfo {author}
  {\bibfnamefont {Y.}~\bibnamefont {Zhang}}, \bibinfo {author} {\bibfnamefont
  {A.}~\bibnamefont {Mahmood}}, \bibinfo {author} {\bibfnamefont
  {Z.}~\bibnamefont {Ouyang}}, \bibinfo {author} {\bibfnamefont
  {H.}~\bibnamefont {Zhang}},\ and\ \bibinfo {author} {\bibfnamefont
  {Z.}~\bibnamefont {Guo}},\ }\bibfield  {title} {\enquote {\bibinfo {title}
  {Recent developments in emerging two-dimensional materials and their
  applications},}\ }\href@noop {} {\bibfield  {journal} {\bibinfo  {journal}
  {Journal of Materials Chemistry C}\ }\textbf {\bibinfo {volume} {8}},\
  \bibinfo {pages} {387--440} (\bibinfo {year} {2020})}\BibitemShut {NoStop}%
\bibitem [{\citenamefont {Chaves}\ \emph {et~al.}(2020)\citenamefont {Chaves},
  \citenamefont {Azadani}, \citenamefont {Alsalman}, \citenamefont {Da~Costa},
  \citenamefont {Frisenda}, \citenamefont {Chaves}, \citenamefont {Song},
  \citenamefont {Kim}, \citenamefont {He}, \citenamefont {Zhou} \emph
  {et~al.}}]{chaves2020bandgap}%
  \BibitemOpen
  \bibfield  {author} {\bibinfo {author} {\bibfnamefont {A.}~\bibnamefont
  {Chaves}}, \bibinfo {author} {\bibfnamefont {J.~G.}\ \bibnamefont {Azadani}},
  \bibinfo {author} {\bibfnamefont {H.}~\bibnamefont {Alsalman}}, \bibinfo
  {author} {\bibfnamefont {D.}~\bibnamefont {Da~Costa}}, \bibinfo {author}
  {\bibfnamefont {R.}~\bibnamefont {Frisenda}}, \bibinfo {author}
  {\bibfnamefont {A.}~\bibnamefont {Chaves}}, \bibinfo {author} {\bibfnamefont
  {S.~H.}\ \bibnamefont {Song}}, \bibinfo {author} {\bibfnamefont {Y.~D.}\
  \bibnamefont {Kim}}, \bibinfo {author} {\bibfnamefont {D.}~\bibnamefont
  {He}}, \bibinfo {author} {\bibfnamefont {J.}~\bibnamefont {Zhou}}, \emph
  {et~al.},\ }\bibfield  {title} {\enquote {\bibinfo {title} {Bandgap
  engineering of two-dimensional semiconductor materials},}\ }\href@noop {}
  {\bibfield  {journal} {\bibinfo  {journal} {npj 2D Materials and
  Applications}\ }\textbf {\bibinfo {volume} {4}},\ \bibinfo {pages} {29}
  (\bibinfo {year} {2020})}\BibitemShut {NoStop}%
\bibitem [{\citenamefont {Xu}\ \emph {et~al.}(2018)\citenamefont {Xu},
  \citenamefont {Liu}, \citenamefont {Sun}, \citenamefont {Cao}, \citenamefont
  {Zhang}, \citenamefont {Wang}, \citenamefont {Liu},\ and\ \citenamefont
  {Liu}}]{xu2018interfacial}%
  \BibitemOpen
  \bibfield  {author} {\bibinfo {author} {\bibfnamefont {X.}~\bibnamefont
  {Xu}}, \bibinfo {author} {\bibfnamefont {C.}~\bibnamefont {Liu}}, \bibinfo
  {author} {\bibfnamefont {Z.}~\bibnamefont {Sun}}, \bibinfo {author}
  {\bibfnamefont {T.}~\bibnamefont {Cao}}, \bibinfo {author} {\bibfnamefont
  {Z.}~\bibnamefont {Zhang}}, \bibinfo {author} {\bibfnamefont
  {E.}~\bibnamefont {Wang}}, \bibinfo {author} {\bibfnamefont {Z.}~\bibnamefont
  {Liu}},\ and\ \bibinfo {author} {\bibfnamefont {K.}~\bibnamefont {Liu}},\
  }\bibfield  {title} {\enquote {\bibinfo {title} {Interfacial engineering in
  graphene bandgap},}\ }\href@noop {} {\bibfield  {journal} {\bibinfo
  {journal} {Chemical Society Reviews}\ }\textbf {\bibinfo {volume} {47}},\
  \bibinfo {pages} {3059--3099} (\bibinfo {year} {2018})}\BibitemShut {NoStop}%
\bibitem [{\citenamefont {Varchon}\ \emph {et~al.}(2007)\citenamefont
  {Varchon}, \citenamefont {Feng}, \citenamefont {Hass}, \citenamefont {Li},
  \citenamefont {Nguyen}, \citenamefont {Naud}, \citenamefont {Mallet},
  \citenamefont {Veuillen}, \citenamefont {Berger}, \citenamefont {Conrad},\
  and\ \citenamefont {Magaud}}]{Varchon2007}%
  \BibitemOpen
  \bibfield  {author} {\bibinfo {author} {\bibfnamefont {F.}~\bibnamefont
  {Varchon}}, \bibinfo {author} {\bibfnamefont {R.}~\bibnamefont {Feng}},
  \bibinfo {author} {\bibfnamefont {J.}~\bibnamefont {Hass}}, \bibinfo {author}
  {\bibfnamefont {X.}~\bibnamefont {Li}}, \bibinfo {author} {\bibfnamefont
  {B.~N.}\ \bibnamefont {Nguyen}}, \bibinfo {author} {\bibfnamefont
  {C.}~\bibnamefont {Naud}}, \bibinfo {author} {\bibfnamefont {P.}~\bibnamefont
  {Mallet}}, \bibinfo {author} {\bibfnamefont {J.-Y.}\ \bibnamefont
  {Veuillen}}, \bibinfo {author} {\bibfnamefont {C.}~\bibnamefont {Berger}},
  \bibinfo {author} {\bibfnamefont {E.~H.}\ \bibnamefont {Conrad}},\ and\
  \bibinfo {author} {\bibfnamefont {L.}~\bibnamefont {Magaud}},\ }\bibfield
  {title} {\enquote {\bibinfo {title} {Electronic structure of epitaxial
  graphene layers on sic: Effect of the substrate},}\ }\href
  {https://doi.org/10.1103/PhysRevLett.99.126805} {\bibfield  {journal}
  {\bibinfo  {journal} {Phys. Rev. Lett.}\ }\textbf {\bibinfo {volume} {99}},\
  \bibinfo {pages} {126805} (\bibinfo {year} {2007})}\BibitemShut {NoStop}%
\bibitem [{\citenamefont {Shemella}\ and\ \citenamefont
  {Nayak}(2009)}]{Shemella2009}%
  \BibitemOpen
  \bibfield  {author} {\bibinfo {author} {\bibfnamefont {P.}~\bibnamefont
  {Shemella}}\ and\ \bibinfo {author} {\bibfnamefont {S.~K.}\ \bibnamefont
  {Nayak}},\ }\bibfield  {title} {\enquote {\bibinfo {title} {Electronic
  structure and band-gap modulation of graphene via substrate surface
  chemistry},}\ }\href {https://doi.org/10.1063/1.3070238} {\bibfield
  {journal} {\bibinfo  {journal} {Applied Physics Letters}\ }\textbf {\bibinfo
  {volume} {94}},\ \bibinfo {pages} {032101} (\bibinfo {year} {2009})},\
  \Eprint {https://arxiv.org/abs/https://doi.org/10.1063/1.3070238}
  {https://doi.org/10.1063/1.3070238} \BibitemShut {NoStop}%
\bibitem [{\citenamefont {Ando}(2015)}]{Ando2015}%
  \BibitemOpen
  \bibfield  {author} {\bibinfo {author} {\bibfnamefont {T.}~\bibnamefont
  {Ando}},\ }\bibfield  {title} {\enquote {\bibinfo {title} {Theory of valley
  hall conductivity in graphene with gap},}\ }\href
  {https://doi.org/10.7566/JPSJ.84.114705} {\bibfield  {journal} {\bibinfo
  {journal} {Journal of the Physical Society of Japan}\ }\textbf {\bibinfo
  {volume} {84}},\ \bibinfo {pages} {114705} (\bibinfo {year} {2015})},\
  \Eprint {https://arxiv.org/abs/https://doi.org/10.7566/JPSJ.84.114705}
  {https://doi.org/10.7566/JPSJ.84.114705} \BibitemShut {NoStop}%
\bibitem [{\citenamefont {Gamayun}\ \emph {et~al.}(2018)\citenamefont
  {Gamayun}, \citenamefont {Ostroukh}, \citenamefont {Gnezdilov}, \citenamefont
  {Adagideli},\ and\ \citenamefont {Beenakker}}]{Gamayun2018}%
  \BibitemOpen
  \bibfield  {author} {\bibinfo {author} {\bibfnamefont {O.~V.}\ \bibnamefont
  {Gamayun}}, \bibinfo {author} {\bibfnamefont {V.~P.}\ \bibnamefont
  {Ostroukh}}, \bibinfo {author} {\bibfnamefont {N.~V.}\ \bibnamefont
  {Gnezdilov}}, \bibinfo {author} {\bibfnamefont {{\.{I}}.}~\bibnamefont
  {Adagideli}},\ and\ \bibinfo {author} {\bibfnamefont {C.~W.~J.}\ \bibnamefont
  {Beenakker}},\ }\bibfield  {title} {\enquote {\bibinfo {title}
  {Valley-momentum locking in a graphene superlattice with y-shaped
  kekul{\'{e}} bond texture},}\ }\href
  {https://doi.org/10.1088/1367-2630/aaa7e5} {\bibfield  {journal} {\bibinfo
  {journal} {New Journal of Physics}\ }\textbf {\bibinfo {volume} {20}},\
  \bibinfo {pages} {023016} (\bibinfo {year} {2018})}\BibitemShut {NoStop}%
\bibitem [{\citenamefont {Mojarro}\ \emph {et~al.}(2020)\citenamefont
  {Mojarro}, \citenamefont {Ibarra-Sierra}, \citenamefont {Sandoval-Santana},
  \citenamefont {Carrillo-Bastos},\ and\ \citenamefont
  {Naumis}}]{Mojarro2020-KekO}%
  \BibitemOpen
  \bibfield  {author} {\bibinfo {author} {\bibfnamefont {M.~A.}\ \bibnamefont
  {Mojarro}}, \bibinfo {author} {\bibfnamefont {V.~G.}\ \bibnamefont
  {Ibarra-Sierra}}, \bibinfo {author} {\bibfnamefont {J.~C.}\ \bibnamefont
  {Sandoval-Santana}}, \bibinfo {author} {\bibfnamefont {R.}~\bibnamefont
  {Carrillo-Bastos}},\ and\ \bibinfo {author} {\bibfnamefont {G.~G.}\
  \bibnamefont {Naumis}},\ }\bibfield  {title} {\enquote {\bibinfo {title}
  {Dynamical floquet spectrum of kekul\'e-distorted graphene under normal
  incidence of electromagnetic radiation},}\ }\href
  {https://doi.org/10.1103/PhysRevB.102.165301} {\bibfield  {journal} {\bibinfo
   {journal} {Phys. Rev. B}\ }\textbf {\bibinfo {volume} {102}},\ \bibinfo
  {pages} {165301} (\bibinfo {year} {2020})}\BibitemShut {NoStop}%
\bibitem [{\citenamefont {Bao}\ \emph {et~al.}(2021)\citenamefont {Bao},
  \citenamefont {Zhang}, \citenamefont {Zhang}, \citenamefont {Wu},
  \citenamefont {Luo}, \citenamefont {Zhou}, \citenamefont {Li}, \citenamefont
  {Hou}, \citenamefont {Yao}, \citenamefont {Liu}, \citenamefont {Yu},
  \citenamefont {Li}, \citenamefont {Duan}, \citenamefont {Yao}, \citenamefont
  {Wang},\ and\ \citenamefont {Zhou}}]{Bao2021}%
  \BibitemOpen
  \bibfield  {author} {\bibinfo {author} {\bibfnamefont {C.}~\bibnamefont
  {Bao}}, \bibinfo {author} {\bibfnamefont {H.}~\bibnamefont {Zhang}}, \bibinfo
  {author} {\bibfnamefont {T.}~\bibnamefont {Zhang}}, \bibinfo {author}
  {\bibfnamefont {X.}~\bibnamefont {Wu}}, \bibinfo {author} {\bibfnamefont
  {L.}~\bibnamefont {Luo}}, \bibinfo {author} {\bibfnamefont {S.}~\bibnamefont
  {Zhou}}, \bibinfo {author} {\bibfnamefont {Q.}~\bibnamefont {Li}}, \bibinfo
  {author} {\bibfnamefont {Y.}~\bibnamefont {Hou}}, \bibinfo {author}
  {\bibfnamefont {W.}~\bibnamefont {Yao}}, \bibinfo {author} {\bibfnamefont
  {L.}~\bibnamefont {Liu}}, \bibinfo {author} {\bibfnamefont {P.}~\bibnamefont
  {Yu}}, \bibinfo {author} {\bibfnamefont {J.}~\bibnamefont {Li}}, \bibinfo
  {author} {\bibfnamefont {W.}~\bibnamefont {Duan}}, \bibinfo {author}
  {\bibfnamefont {H.}~\bibnamefont {Yao}}, \bibinfo {author} {\bibfnamefont
  {Y.}~\bibnamefont {Wang}},\ and\ \bibinfo {author} {\bibfnamefont
  {S.}~\bibnamefont {Zhou}},\ }\bibfield  {title} {\enquote {\bibinfo {title}
  {Experimental evidence of chiral symmetry breaking in kekul\'e-ordered
  graphene},}\ }\href {https://doi.org/10.1103/PhysRevLett.126.206804}
  {\bibfield  {journal} {\bibinfo  {journal} {Phys. Rev. Lett.}\ }\textbf
  {\bibinfo {volume} {126}},\ \bibinfo {pages} {206804} (\bibinfo {year}
  {2021})}\BibitemShut {NoStop}%
\bibitem [{\citenamefont {Bao}\ \emph {et~al.}(2022)\citenamefont {Bao},
  \citenamefont {Zhang}, \citenamefont {Wu}, \citenamefont {Zhou},
  \citenamefont {Li}, \citenamefont {Yu}, \citenamefont {Li}, \citenamefont
  {Duan},\ and\ \citenamefont {Zhou}}]{Bao2022}%
  \BibitemOpen
  \bibfield  {author} {\bibinfo {author} {\bibfnamefont {C.}~\bibnamefont
  {Bao}}, \bibinfo {author} {\bibfnamefont {H.}~\bibnamefont {Zhang}}, \bibinfo
  {author} {\bibfnamefont {X.}~\bibnamefont {Wu}}, \bibinfo {author}
  {\bibfnamefont {S.}~\bibnamefont {Zhou}}, \bibinfo {author} {\bibfnamefont
  {Q.}~\bibnamefont {Li}}, \bibinfo {author} {\bibfnamefont {P.}~\bibnamefont
  {Yu}}, \bibinfo {author} {\bibfnamefont {J.}~\bibnamefont {Li}}, \bibinfo
  {author} {\bibfnamefont {W.}~\bibnamefont {Duan}},\ and\ \bibinfo {author}
  {\bibfnamefont {S.}~\bibnamefont {Zhou}},\ }\bibfield  {title} {\enquote
  {\bibinfo {title} {Coexistence of extended flat band and kekul\'e order in
  li-intercalated graphene},}\ }\href
  {https://doi.org/10.1103/PhysRevB.105.L161106} {\bibfield  {journal}
  {\bibinfo  {journal} {Phys. Rev. B}\ }\textbf {\bibinfo {volume} {105}},\
  \bibinfo {pages} {L161106} (\bibinfo {year} {2022})}\BibitemShut {NoStop}%
\bibitem [{\citenamefont {Eom}\ and\ \citenamefont {Koo}(2020)}]{EOM2022}%
  \BibitemOpen
  \bibfield  {author} {\bibinfo {author} {\bibfnamefont {D.}~\bibnamefont
  {Eom}}\ and\ \bibinfo {author} {\bibfnamefont {J.-Y.}\ \bibnamefont {Koo}},\
  }\bibfield  {title} {\enquote {\bibinfo {title} {Direct measurement of
  strain-driven kekulé distortion in graphene and its electronic
  properties},}\ }\href {https://doi.org/10.1039/D0NR03565C} {\bibfield
  {journal} {\bibinfo  {journal} {Nanoscale}\ }\textbf {\bibinfo {volume}
  {12}},\ \bibinfo {pages} {19604--19608} (\bibinfo {year} {2020})}\BibitemShut
  {NoStop}%
\bibitem [{\citenamefont {Qu}\ \emph {et~al.}(2022)\citenamefont {Qu},
  \citenamefont {Nigge}, \citenamefont {Link}, \citenamefont {Levy},
  \citenamefont {Michiardi}, \citenamefont {Spandar}, \citenamefont {Matthé},
  \citenamefont {Schneider}, \citenamefont {Zhdanovich}, \citenamefont
  {Starke}, \citenamefont {Gutiérrez},\ and\ \citenamefont
  {Damascelli}}]{Qu2022}%
  \BibitemOpen
  \bibfield  {author} {\bibinfo {author} {\bibfnamefont {A.~C.}\ \bibnamefont
  {Qu}}, \bibinfo {author} {\bibfnamefont {P.}~\bibnamefont {Nigge}}, \bibinfo
  {author} {\bibfnamefont {S.}~\bibnamefont {Link}}, \bibinfo {author}
  {\bibfnamefont {G.}~\bibnamefont {Levy}}, \bibinfo {author} {\bibfnamefont
  {M.}~\bibnamefont {Michiardi}}, \bibinfo {author} {\bibfnamefont {P.~L.}\
  \bibnamefont {Spandar}}, \bibinfo {author} {\bibfnamefont {T.}~\bibnamefont
  {Matthé}}, \bibinfo {author} {\bibfnamefont {M.}~\bibnamefont {Schneider}},
  \bibinfo {author} {\bibfnamefont {S.}~\bibnamefont {Zhdanovich}}, \bibinfo
  {author} {\bibfnamefont {U.}~\bibnamefont {Starke}}, \bibinfo {author}
  {\bibfnamefont {C.}~\bibnamefont {Gutiérrez}},\ and\ \bibinfo {author}
  {\bibfnamefont {A.}~\bibnamefont {Damascelli}},\ }\bibfield  {title}
  {\enquote {\bibinfo {title} {Ubiquitous defect-induced density wave
  instability in monolayer graphene},}\ }\href
  {https://doi.org/10.1126/sciadv.abm5180} {\bibfield  {journal} {\bibinfo
  {journal} {Science Advances}\ }\textbf {\bibinfo {volume} {8}},\ \bibinfo
  {pages} {eabm5180} (\bibinfo {year} {2022})},\ \Eprint
  {https://arxiv.org/abs/https://www.science.org/doi/pdf/10.1126/sciadv.abm5180}
  {https://www.science.org/doi/pdf/10.1126/sciadv.abm5180} \BibitemShut
  {NoStop}%
\bibitem [{\citenamefont {Andrade}\ \emph {et~al.}(2022)\citenamefont
  {Andrade}, \citenamefont {Carrillo-Bastos}, \citenamefont {Asmar},\ and\
  \citenamefont {Naumis}}]{KekuleAndrade2022}%
  \BibitemOpen
  \bibfield  {author} {\bibinfo {author} {\bibfnamefont {E.}~\bibnamefont
  {Andrade}}, \bibinfo {author} {\bibfnamefont {R.}~\bibnamefont
  {Carrillo-Bastos}}, \bibinfo {author} {\bibfnamefont {M.~M.}\ \bibnamefont
  {Asmar}},\ and\ \bibinfo {author} {\bibfnamefont {G.~G.}\ \bibnamefont
  {Naumis}},\ }\bibfield  {title} {\enquote {\bibinfo {title} {Kekul\'e-induced
  valley birefringence and skew scattering in graphene},}\ }\href
  {https://doi.org/10.1103/PhysRevB.106.195413} {\bibfield  {journal} {\bibinfo
   {journal} {Phys. Rev. B}\ }\textbf {\bibinfo {volume} {106}},\ \bibinfo
  {pages} {195413} (\bibinfo {year} {2022})}\BibitemShut {NoStop}%
\bibitem [{\citenamefont {Zeng}\ and\ \citenamefont {Shen}(2022)}]{Zeng2022}%
  \BibitemOpen
  \bibfield  {author} {\bibinfo {author} {\bibfnamefont {W.}~\bibnamefont
  {Zeng}}\ and\ \bibinfo {author} {\bibfnamefont {R.}~\bibnamefont {Shen}},\
  }\bibfield  {title} {\enquote {\bibinfo {title} {Light-modulated josephson
  effect in kekul\'e patterned graphene},}\ }\href
  {https://doi.org/10.1103/PhysRevB.105.094510} {\bibfield  {journal} {\bibinfo
   {journal} {Phys. Rev. B}\ }\textbf {\bibinfo {volume} {105}},\ \bibinfo
  {pages} {094510} (\bibinfo {year} {2022})}\BibitemShut {NoStop}%
\bibitem [{\citenamefont {Garc\'{\i}a}, \citenamefont {Stegmann},\ and\
  \citenamefont {Betancur-Ocampo}(2022)}]{Stegmann2022}%
  \BibitemOpen
  \bibfield  {author} {\bibinfo {author} {\bibfnamefont {S.~G.~y.}\
  \bibnamefont {Garc\'{\i}a}}, \bibinfo {author} {\bibfnamefont
  {T.}~\bibnamefont {Stegmann}},\ and\ \bibinfo {author} {\bibfnamefont
  {Y.}~\bibnamefont {Betancur-Ocampo}},\ }\bibfield  {title} {\enquote
  {\bibinfo {title} {Generalized hamiltonian for kekul\'e graphene and the
  emergence of valley-cooperative klein tunneling},}\ }\href
  {https://doi.org/10.1103/PhysRevB.105.125139} {\bibfield  {journal} {\bibinfo
   {journal} {Phys. Rev. B}\ }\textbf {\bibinfo {volume} {105}},\ \bibinfo
  {pages} {125139} (\bibinfo {year} {2022})}\BibitemShut {NoStop}%
\bibitem [{\citenamefont {Hartmann}, \citenamefont {Robinson},\ and\
  \citenamefont {Portnoi}(2010)}]{Hartmann2010Waveguides}%
  \BibitemOpen
  \bibfield  {author} {\bibinfo {author} {\bibfnamefont {R.~R.}\ \bibnamefont
  {Hartmann}}, \bibinfo {author} {\bibfnamefont {N.~J.}\ \bibnamefont
  {Robinson}},\ and\ \bibinfo {author} {\bibfnamefont {M.~E.}\ \bibnamefont
  {Portnoi}},\ }\bibfield  {title} {\enquote {\bibinfo {title} {Smooth electron
  waveguides in graphene},}\ }\href
  {https://doi.org/10.1103/PhysRevB.81.245431} {\bibfield  {journal} {\bibinfo
  {journal} {Phys. Rev. B}\ }\textbf {\bibinfo {volume} {81}},\ \bibinfo
  {pages} {245431} (\bibinfo {year} {2010})}\BibitemShut {NoStop}%
\bibitem [{\citenamefont {Hartmann}\ and\ \citenamefont
  {Portnoi}(2017)}]{Hartmann2017}%
  \BibitemOpen
  \bibfield  {author} {\bibinfo {author} {\bibfnamefont {R.~R.}\ \bibnamefont
  {Hartmann}}\ and\ \bibinfo {author} {\bibfnamefont {M.~E.}\ \bibnamefont
  {Portnoi}},\ }\bibfield  {title} {\enquote {\bibinfo {title} {Two-dimensional
  dirac particles in a p{\"o}schl-teller waveguide},}\ }\href
  {https://doi.org/10.1038/s41598-017-11411-w} {\bibfield  {journal} {\bibinfo
  {journal} {Scientific Reports}\ }\textbf {\bibinfo {volume} {7}},\ \bibinfo
  {pages} {11599} (\bibinfo {year} {2017})}\BibitemShut {NoStop}%
\bibitem [{\citenamefont {Hartmann}\ and\ \citenamefont
  {Portnoi}(2020)}]{Hartmann2020Waveguides}%
  \BibitemOpen
  \bibfield  {author} {\bibinfo {author} {\bibfnamefont {R.~R.}\ \bibnamefont
  {Hartmann}}\ and\ \bibinfo {author} {\bibfnamefont {M.~E.}\ \bibnamefont
  {Portnoi}},\ }\bibfield  {title} {\enquote {\bibinfo {title} {Bipolar
  electron waveguides in graphene},}\ }\href
  {https://doi.org/10.1103/PhysRevB.102.155421} {\bibfield  {journal} {\bibinfo
   {journal} {Phys. Rev. B}\ }\textbf {\bibinfo {volume} {102}},\ \bibinfo
  {pages} {155421} (\bibinfo {year} {2020})}\BibitemShut {NoStop}%
\bibitem [{\citenamefont {Carrillo-Bastos}\ \emph {et~al.}(2014)\citenamefont
  {Carrillo-Bastos}, \citenamefont {Faria}, \citenamefont {Latg\'e},
  \citenamefont {Mireles},\ and\ \citenamefont
  {Sandler}}]{Carrillo2014Gaussian}%
  \BibitemOpen
  \bibfield  {author} {\bibinfo {author} {\bibfnamefont {R.}~\bibnamefont
  {Carrillo-Bastos}}, \bibinfo {author} {\bibfnamefont {D.}~\bibnamefont
  {Faria}}, \bibinfo {author} {\bibfnamefont {A.}~\bibnamefont {Latg\'e}},
  \bibinfo {author} {\bibfnamefont {F.}~\bibnamefont {Mireles}},\ and\ \bibinfo
  {author} {\bibfnamefont {N.}~\bibnamefont {Sandler}},\ }\bibfield  {title}
  {\enquote {\bibinfo {title} {Gaussian deformations in graphene ribbons:
  Flowers and confinement},}\ }\href
  {https://doi.org/10.1103/PhysRevB.90.041411} {\bibfield  {journal} {\bibinfo
  {journal} {Phys. Rev. B}\ }\textbf {\bibinfo {volume} {90}},\ \bibinfo
  {pages} {041411} (\bibinfo {year} {2014})}\BibitemShut {NoStop}%
\bibitem [{\citenamefont {Cao}\ \emph {et~al.}(2017)\citenamefont {Cao},
  \citenamefont {Zhou}, \citenamefont {Wei},\ and\ \citenamefont
  {Cheng}}]{Cao2017Waveguides}%
  \BibitemOpen
  \bibfield  {author} {\bibinfo {author} {\bibfnamefont {S.-M.}\ \bibnamefont
  {Cao}}, \bibinfo {author} {\bibfnamefont {J.-J.}\ \bibnamefont {Zhou}},
  \bibinfo {author} {\bibfnamefont {X.}~\bibnamefont {Wei}},\ and\ \bibinfo
  {author} {\bibfnamefont {S.-G.}\ \bibnamefont {Cheng}},\ }\bibfield  {title}
  {\enquote {\bibinfo {title} {Investigation of valley-resolved transmission
  through gate defined graphene carrier guiders},}\ }\href
  {https://doi.org/10.1088/1361-648x/aa5a7f} {\bibfield  {journal} {\bibinfo
  {journal} {Journal of Physics: Condensed Matter}\ }\textbf {\bibinfo {volume}
  {29}},\ \bibinfo {pages} {145301} (\bibinfo {year} {2017})}\BibitemShut
  {NoStop}%
\bibitem [{\citenamefont {Mosallanejad}\ \emph {et~al.}(2018)\citenamefont
  {Mosallanejad}, \citenamefont {Wang}, \citenamefont {Qiao},\ and\
  \citenamefont {Guo}}]{Mosallanejad2018Waveguides}%
  \BibitemOpen
  \bibfield  {author} {\bibinfo {author} {\bibfnamefont {V.}~\bibnamefont
  {Mosallanejad}}, \bibinfo {author} {\bibfnamefont {K.}~\bibnamefont {Wang}},
  \bibinfo {author} {\bibfnamefont {Z.}~\bibnamefont {Qiao}},\ and\ \bibinfo
  {author} {\bibfnamefont {G.}~\bibnamefont {Guo}},\ }\bibfield  {title}
  {\enquote {\bibinfo {title} {Perfectly conducting graphene electronic
  waveguide with curved channels},}\ }\href
  {https://doi.org/10.1088/1361-648x/aacfca} {\bibfield  {journal} {\bibinfo
  {journal} {Journal of Physics: Condensed Matter}\ }\textbf {\bibinfo {volume}
  {30}},\ \bibinfo {pages} {325301} (\bibinfo {year} {2018})}\BibitemShut
  {NoStop}%
\bibitem [{\citenamefont {Robles-Raygoza}\ \emph {et~al.}(2022)\citenamefont
  {Robles-Raygoza}, \citenamefont {Ibarra-Sierra}, \citenamefont
  {Sandoval-Santana},\ and\ \citenamefont {Carrillo-Bastos}}]{Raygoza2022}%
  \BibitemOpen
  \bibfield  {author} {\bibinfo {author} {\bibfnamefont {E.~J.}\ \bibnamefont
  {Robles-Raygoza}}, \bibinfo {author} {\bibfnamefont {V.~G.}\ \bibnamefont
  {Ibarra-Sierra}}, \bibinfo {author} {\bibfnamefont {J.~C.}\ \bibnamefont
  {Sandoval-Santana}},\ and\ \bibinfo {author} {\bibfnamefont {R.}~\bibnamefont
  {Carrillo-Bastos}},\ }\bibfield  {title} {\enquote {\bibinfo {title}
  {Ballistic guided electrons against disorder in graphene nanoribbons},}\
  }\href {https://doi.org/10.1063/5.0116479} {\bibfield  {journal} {\bibinfo
  {journal} {Journal of Applied Physics}\ }\textbf {\bibinfo {volume} {132}},\
  \bibinfo {pages} {164305} (\bibinfo {year} {2022})},\ \Eprint
  {https://arxiv.org/abs/https://doi.org/10.1063/5.0116479}
  {https://doi.org/10.1063/5.0116479} \BibitemShut {NoStop}%
\bibitem [{\citenamefont {Ng}\ \emph {et~al.}(2021)\citenamefont {Ng},
  \citenamefont {Wild}, \citenamefont {Portnoi},\ and\ \citenamefont
  {Hartmann}}]{Hartmann2021}%
  \BibitemOpen
  \bibfield  {author} {\bibinfo {author} {\bibfnamefont {R.~A.}\ \bibnamefont
  {Ng}}, \bibinfo {author} {\bibfnamefont {A.}~\bibnamefont {Wild}}, \bibinfo
  {author} {\bibfnamefont {M.~E.}\ \bibnamefont {Portnoi}},\ and\ \bibinfo
  {author} {\bibfnamefont {R.~R.}\ \bibnamefont {Hartmann}},\ }\href@noop {}
  {\enquote {\bibinfo {title} {Mapping borophene onto graphene: Quasi-exact
  solutions for guiding potentials in tilted dirac cones},}\ } (\bibinfo {year}
  {2021}),\ \Eprint {https://arxiv.org/abs/2111.10760} {arXiv:2111.10760
  [cond-mat.mes-hall]} \BibitemShut {NoStop}%
\bibitem [{\citenamefont {Addanki}, \citenamefont {Amiri},\ and\ \citenamefont
  {Yupapin}(2018)}]{Addanki2018Review}%
  \BibitemOpen
  \bibfield  {author} {\bibinfo {author} {\bibfnamefont {S.}~\bibnamefont
  {Addanki}}, \bibinfo {author} {\bibfnamefont {I.}~\bibnamefont {Amiri}},\
  and\ \bibinfo {author} {\bibfnamefont {P.}~\bibnamefont {Yupapin}},\
  }\bibfield  {title} {\enquote {\bibinfo {title} {Review of optical
  fibers-introduction and applications in fiber lasers},}\ }\href
  {https://doi.org/https://doi.org/10.1016/j.rinp.2018.07.028} {\bibfield
  {journal} {\bibinfo  {journal} {Results in Physics}\ }\textbf {\bibinfo
  {volume} {10}},\ \bibinfo {pages} {743--750} (\bibinfo {year}
  {2018})}\BibitemShut {NoStop}%
\bibitem [{\citenamefont {Dragic}, \citenamefont {Cavillon},\ and\
  \citenamefont {Ballato}(2018)}]{Dragic2018}%
  \BibitemOpen
  \bibfield  {author} {\bibinfo {author} {\bibfnamefont {P.~D.}\ \bibnamefont
  {Dragic}}, \bibinfo {author} {\bibfnamefont {M.}~\bibnamefont {Cavillon}},\
  and\ \bibinfo {author} {\bibfnamefont {J.}~\bibnamefont {Ballato}},\
  }\bibfield  {title} {\enquote {\bibinfo {title} {Materials for optical fiber
  lasers: A review},}\ }\href {https://doi.org/10.1063/1.5048410} {\bibfield
  {journal} {\bibinfo  {journal} {Applied Physics Reviews}\ }\textbf {\bibinfo
  {volume} {5}},\ \bibinfo {pages} {041301} (\bibinfo {year} {2018})},\ \Eprint
  {https://arxiv.org/abs/https://doi.org/10.1063/1.5048410}
  {https://doi.org/10.1063/1.5048410} \BibitemShut {NoStop}%
\bibitem [{\citenamefont {Carrillo-Bastos}\ \emph {et~al.}(2016)\citenamefont
  {Carrillo-Bastos}, \citenamefont {Le{\'o}n}, \citenamefont {Faria},
  \citenamefont {Latg{\'e}}, \citenamefont {Andrei},\ and\ \citenamefont
  {Sandler}}]{Carrillo2016strained}%
  \BibitemOpen
  \bibfield  {author} {\bibinfo {author} {\bibfnamefont {R.}~\bibnamefont
  {Carrillo-Bastos}}, \bibinfo {author} {\bibfnamefont {C.}~\bibnamefont
  {Le{\'o}n}}, \bibinfo {author} {\bibfnamefont {D.}~\bibnamefont {Faria}},
  \bibinfo {author} {\bibfnamefont {A.}~\bibnamefont {Latg{\'e}}}, \bibinfo
  {author} {\bibfnamefont {E.~Y.}\ \bibnamefont {Andrei}},\ and\ \bibinfo
  {author} {\bibfnamefont {N.}~\bibnamefont {Sandler}},\ }\bibfield  {title}
  {\enquote {\bibinfo {title} {Strained fold-assisted transport in graphene
  systems},}\ }\href@noop {} {\bibfield  {journal} {\bibinfo  {journal}
  {Physical Review B}\ }\textbf {\bibinfo {volume} {94}},\ \bibinfo {pages}
  {125422} (\bibinfo {year} {2016})}\BibitemShut {NoStop}%
\bibitem [{\citenamefont {Giambastiani}\ \emph {et~al.}(2022)\citenamefont
  {Giambastiani}, \citenamefont {Tommasi}, \citenamefont {Bianco},
  \citenamefont {Fabbri}, \citenamefont {Coletti}, \citenamefont {Tredicucci},
  \citenamefont {Pitanti},\ and\ \citenamefont {Roddaro}}]{Giambastiani2022}%
  \BibitemOpen
  \bibfield  {author} {\bibinfo {author} {\bibfnamefont {D.}~\bibnamefont
  {Giambastiani}}, \bibinfo {author} {\bibfnamefont {C.}~\bibnamefont
  {Tommasi}}, \bibinfo {author} {\bibfnamefont {F.}~\bibnamefont {Bianco}},
  \bibinfo {author} {\bibfnamefont {F.}~\bibnamefont {Fabbri}}, \bibinfo
  {author} {\bibfnamefont {C.}~\bibnamefont {Coletti}}, \bibinfo {author}
  {\bibfnamefont {A.}~\bibnamefont {Tredicucci}}, \bibinfo {author}
  {\bibfnamefont {A.}~\bibnamefont {Pitanti}},\ and\ \bibinfo {author}
  {\bibfnamefont {S.}~\bibnamefont {Roddaro}},\ }\bibfield  {title} {\enquote
  {\bibinfo {title} {Strain-engineered wrinkles on graphene using polymeric
  actuators},}\ }\href {https://doi.org/10.1103/PhysRevApplied.18.024069}
  {\bibfield  {journal} {\bibinfo  {journal} {Phys. Rev. Appl.}\ }\textbf
  {\bibinfo {volume} {18}},\ \bibinfo {pages} {024069} (\bibinfo {year}
  {2022})}\BibitemShut {NoStop}%
\bibitem [{\citenamefont {Mucciolo}, \citenamefont {Castro~Neto},\ and\
  \citenamefont {Lewenkopf}(2009)}]{Mucciolo2009}%
  \BibitemOpen
  \bibfield  {author} {\bibinfo {author} {\bibfnamefont {E.~R.}\ \bibnamefont
  {Mucciolo}}, \bibinfo {author} {\bibfnamefont {A.~H.}\ \bibnamefont
  {Castro~Neto}},\ and\ \bibinfo {author} {\bibfnamefont {C.~H.}\ \bibnamefont
  {Lewenkopf}},\ }\bibfield  {title} {\enquote {\bibinfo {title} {Conductance
  quantization and transport gaps in disordered graphene nanoribbons},}\ }\href
  {https://doi.org/10.1103/PhysRevB.79.075407} {\bibfield  {journal} {\bibinfo
  {journal} {Phys. Rev. B}\ }\textbf {\bibinfo {volume} {79}},\ \bibinfo
  {pages} {075407} (\bibinfo {year} {2009})}\BibitemShut {NoStop}%
\bibitem [{\citenamefont {Maier}(2007)}]{Maier2007}%
  \BibitemOpen
  \bibfield  {author} {\bibinfo {author} {\bibfnamefont {R.}~\bibnamefont
  {Maier}},\ }\bibfield  {title} {\enquote {\bibinfo {title} {The 192 solutions
  of the heun equation},}\ }\href@noop {} {\bibfield  {journal} {\bibinfo
  {journal} {Mathematics of Computation}\ }\textbf {\bibinfo {volume} {76}},\
  \bibinfo {pages} {811--843} (\bibinfo {year} {2007})}\BibitemShut {NoStop}%
\bibitem [{\citenamefont {Olver}\ \emph {et~al.}(2010)\citenamefont {Olver},
  \citenamefont {Lozier}, \citenamefont {Boisvert},\ and\ \citenamefont
  {Clark}}]{Olver2010nist}%
  \BibitemOpen
  \bibfield  {author} {\bibinfo {author} {\bibfnamefont {F.~W.}\ \bibnamefont
  {Olver}}, \bibinfo {author} {\bibfnamefont {D.~W.}\ \bibnamefont {Lozier}},
  \bibinfo {author} {\bibfnamefont {R.~F.}\ \bibnamefont {Boisvert}},\ and\
  \bibinfo {author} {\bibfnamefont {C.~W.}\ \bibnamefont {Clark}},\ }\href@noop
  {} {\emph {\bibinfo {title} {NIST handbook of mathematical functions hardback
  and CD-ROM}}}\ (\bibinfo  {publisher} {Cambridge university press},\ \bibinfo
  {year} {2010})\BibitemShut {NoStop}%
\bibitem [{\citenamefont {Downing}(2013)}]{Downing2013}%
  \BibitemOpen
  \bibfield  {author} {\bibinfo {author} {\bibfnamefont {C.~A.}\ \bibnamefont
  {Downing}},\ }\bibfield  {title} {\enquote {\bibinfo {title} {On a solution
  of the schrödinger equation with a hyperbolic double-well potential},}\
  }\href {https://doi.org/10.1063/1.4811855} {\bibfield  {journal} {\bibinfo
  {journal} {Journal of Mathematical Physics}\ }\textbf {\bibinfo {volume}
  {54}},\ \bibinfo {pages} {072101} (\bibinfo {year} {2013})},\ \Eprint
  {https://arxiv.org/abs/https://doi.org/10.1063/1.4811855}
  {https://doi.org/10.1063/1.4811855} \BibitemShut {NoStop}%
\bibitem [{\citenamefont {Beenakker}\ \emph {et~al.}(2018)\citenamefont
  {Beenakker}, \citenamefont {Gnezdilov}, \citenamefont {Dresselhaus},
  \citenamefont {Ostroukh}, \citenamefont {Herasymenko}, \citenamefont
  {Adagideli},\ and\ \citenamefont {Tworzyd\l{}o}}]{Beenakker2018}%
  \BibitemOpen
  \bibfield  {author} {\bibinfo {author} {\bibfnamefont {C.~W.~J.}\
  \bibnamefont {Beenakker}}, \bibinfo {author} {\bibfnamefont {N.~V.}\
  \bibnamefont {Gnezdilov}}, \bibinfo {author} {\bibfnamefont {E.}~\bibnamefont
  {Dresselhaus}}, \bibinfo {author} {\bibfnamefont {V.~P.}\ \bibnamefont
  {Ostroukh}}, \bibinfo {author} {\bibfnamefont {Y.}~\bibnamefont
  {Herasymenko}}, \bibinfo {author} {\bibfnamefont {i.~d.~I.}\ \bibnamefont
  {Adagideli}},\ and\ \bibinfo {author} {\bibfnamefont {J.}~\bibnamefont
  {Tworzyd\l{}o}},\ }\bibfield  {title} {\enquote {\bibinfo {title} {Valley
  switch in a graphene superlattice due to pseudo-andreev reflection},}\ }\href
  {https://doi.org/10.1103/PhysRevB.97.241403} {\bibfield  {journal} {\bibinfo
  {journal} {Phys. Rev. B}\ }\textbf {\bibinfo {volume} {97}},\ \bibinfo
  {pages} {241403} (\bibinfo {year} {2018})}\BibitemShut {NoStop}%
\bibitem [{\citenamefont {Henderson}\ and\ \citenamefont
  {Searle}(1981)}]{Harold1981Kronecker}%
  \BibitemOpen
  \bibfield  {author} {\bibinfo {author} {\bibfnamefont {H.~V.}\ \bibnamefont
  {Henderson}}\ and\ \bibinfo {author} {\bibfnamefont {S.~R.}\ \bibnamefont
  {Searle}},\ }\bibfield  {title} {\enquote {\bibinfo {title} {The
  vec-permutation matrix, the vec operator and kronecker products: a review},}\
  }\href {https://doi.org/10.1080/03081088108817379} {\bibfield  {journal}
  {\bibinfo  {journal} {Linear and Multilinear Algebra}\ }\textbf {\bibinfo
  {volume} {9}},\ \bibinfo {pages} {271--288} (\bibinfo {year} {1981})},\
  \Eprint {https://arxiv.org/abs/https://doi.org/10.1080/03081088108817379}
  {https://doi.org/10.1080/03081088108817379} \BibitemShut {NoStop}%
\bibitem [{\citenamefont {Loan}(2000)}]{Loan2000Kronecker}%
  \BibitemOpen
  \bibfield  {author} {\bibinfo {author} {\bibfnamefont {C.~F.}\ \bibnamefont
  {Loan}},\ }\bibfield  {title} {\enquote {\bibinfo {title} {The ubiquitous
  kronecker product},}\ }\href
  {https://doi.org/https://doi.org/10.1016/S0377-0427(00)00393-9} {\bibfield
  {journal} {\bibinfo  {journal} {Journal of Computational and Applied
  Mathematics}\ }\textbf {\bibinfo {volume} {123}},\ \bibinfo {pages} {85--100}
  (\bibinfo {year} {2000})},\ \bibinfo {note} {numerical Analysis 2000. Vol.
  III: Linear Algebra}\BibitemShut {NoStop}%
\bibitem [{\citenamefont {Shankar}(2012)}]{shankar2012principles}%
  \BibitemOpen
  \bibfield  {author} {\bibinfo {author} {\bibfnamefont {R.}~\bibnamefont
  {Shankar}},\ }\href@noop {} {\emph {\bibinfo {title} {Principles of quantum
  mechanics}}}\ (\bibinfo  {publisher} {Springer Science \& Business Media},\
  \bibinfo {year} {2012})\BibitemShut {NoStop}%
\end{thebibliography}%

\end{document}